\documentclass[numberedappendix, iop, apj]{emulateapj}
\usepackage{natbib}
\usepackage{mathtools}
\usepackage{hyperref}

\def\lya{Ly-$\alpha$}
\def\nh{$N_{\text{H\,\tiny{I}}}$}

\def\elc{$\ell_{\rm{c}}$}
\def\w1526{$W_{1526}$}

\def\kms{km~s$^{-1}$}

\def\hi{\ion{H}{1}}
\def\siii{\ion{Si}{2}}
\def\cii{\ion{C}{2}}
\def\siiistar{\ion{Si}{2}$^*$}
\def\ciistar{\ion{C}{2}$^*$}

\def\lambdasiiistar{\ion{Si}{2}$^*$ $\lambda$1264.7}
\def\lambdaciistar{\ion{C}{2}$^*$ $\lambda$1335.7}
\def\lambdasiiistartwo{\ion{Si}{2}$^*$ $\lambda$1533.4}
\def\lambdasiiistarthree{\ion{Si}{2}$^*$ $\lambda$1265.0}
\def\lambdasiii{\ion{Si}{2} $\lambda$1808.0}
\def\nhi{n$_{\rm{H}}$}
\def\nel{n$_{\rm{e}}$}
\def\rsi{$r_{\rm{Si}}$}
\def\rc{$r_{\rm{C}}$}

\shorttitle{The Physical Conditions of High-$z$ Atomic Gas}
\shortauthors{Neeleman et al.}

\begin{document}

\title{Probing the Physical Conditions of Atomic Gas at High Redshift}

\author{Marcel Neeleman}
\affil{Department of Physics and Center for Astrophysics and Space Sciences, UCSD, La Jolla, CA 92093, USA}
\email{mneeleman@physics.ucsd.edu}
\author{J. Xavier Prochaska}
\affil{Department of Astronomy \& Astrophysics, UCO/Lick Observatory, 1156 High Street, \\
University of California, Santa Cruz, CA 95064, USA}
\and
\author{Arthur M. Wolfe}
\affil{Department of Physics and Center for Astrophysics and Space Sciences, UCSD, La Jolla, CA 92093, USA}

\begin{abstract} 
A new method is used to measure the physical conditions of the gas in damped Lyman-$\alpha$ systems (DLAs). Using high resolution absorption spectra of a sample of 80 DLAs, we are able to measure the ratio of the upper and lower fine-structure levels of the ground state of C$^+$ and Si$^+$. These ratios are determined solely by the physical conditions of the gas. We explore the allowed physical parameter space using a Monte Carlo Markov Chain method to constrain simultaneously the temperature, neutral hydrogen density, and electron density of each DLA. The results indicate that at least 5~\% of all DLAs have the bulk of their gas in a dense, cold phase with typical densities of $\sim$ 100 cm$^{-3}$ and temperatures below 500 K. We further find that the typical pressure of DLAs in our sample is $\log(P/k_{\rm{B}})$ = 3.4 [K cm$^{-3}$], which is comparable to the pressure of the local interstellar medium (ISM), and that the components containing the bulk of the neutral gas can be quite small with absorption sizes as small as a few parsec. We show that the majority of the systems are consistent with having densities significantly higher than expected from a purely canonical WNM, indicating that significant quantities of dense gas (i.e. {\nhi} $>$ 0.1 cm$^{-3}$) are required to match observations. Finally, we identify 8 systems with positive detections of {\siiistar}. These systems have pressures ($P/k_{\rm{B}}$) in excess of 20000 K cm$^{-3}$, which suggest that these systems tag a highly turbulent ISM in young, star-forming galaxies.\\

\end{abstract}


\section{Introduction}
\label{sec:intro}

Galaxy formation and evolution is fundamentally dependent on the gas that forms the galaxy. From its inception as a gravitational bound collection of gas to the formation of stars inside an evolved galaxy, the physical properties of the gas affect the outcome of such processes. It is therefore of paramount importance to understand the physical properties of the gas inside and around galaxies.

Already early on, \citet{Field1969} noted that when heating and cooling sources of a neutral gas are in thermal equilibrium, the gas naturally segregates into two distinct phases. A cold, dense phase known as the cold neutral medium (CNM), and a warmer, less dense phase, the warm neutral medium (WNM). This model was improved upon by \citet{McKee1977} to include a third phase, namely the hot ionized medium due to ionizing shock fronts produced by supernova. On the basis of this model, \citet{Wolfire1995,Wolfire2003} calculated the equilibrium pressures and temperatures of the neutral gas under a variety of different galactic conditions. This theoretical model, in its general form, still holds as the paradigm for the physical conditions of neutral galactic gas.

\begin{deluxetable*}{llccccccccl}
\tabletypesize{\scriptsize}
%\rotate
\tablecaption{{Fine Structure DLA sample}
\label{tab:meas}}
\tablewidth{0pt}
\tablehead{
\colhead{Index} &
\colhead{QSO} &
\colhead{$z_{\rm{abs}}$} &
\colhead{$\log N_{\rm HI}$} &
\colhead{Metallicity} &
\colhead{M\,\tablenotemark{a}} &
\colhead{${\log} N_{\rm CII*}$} &
\colhead{${\log} N_{\rm SiII*}$} &
\colhead{${\log} N_{\rm SiII}$} &
\colhead{N13\,\tablenotemark{b}} &
\colhead{References} \\
\colhead{Number} &
\colhead{} &
\colhead{} &
\colhead{(cm$^{-2}$)} &
\colhead{[M/H]} &
\colhead{} &
\colhead{(cm$^{-2}$)} &
\colhead{(cm$^{-2}$)} &
\colhead{(cm$^{-2}$)} &
\colhead{} &
\colhead{} \\
}
\startdata
1 & Q1157$+$014 & 1.9437 & 21.70 $\pm$ 0.10 & $-1.23$ $\pm$ 0.10 & Si & $>$ $14.75$\,\tablenotemark{c} & $12.37$$\pm$0.03 & $15.97$$\pm$0.01 & Y & 5, 18\\
2 & Q1215$+$33 & 1.9991 & 20.95 $\pm$ 0.06 & $-1.43$ $\pm$ 0.07 & Si & $<$ $13.18$ & $<$ $12.56$ & $15.02$$\pm$0.02 & Y & 4, 7\\
3 & Q0458$-$02 & 2.0395 & 21.65 $\pm$ 0.09 & $-1.11$ $\pm$ 0.09 & Si & $>$ $14.88$\,\tablenotemark{c} & $<$ $12.58$ & $16.04$$\pm$0.02\,\tablenotemark{c} & Y & 4, 7\\
4 & J2340$-$0053 & 2.0545 & 20.35 $\pm$ 0.15 & $-0.55$ $\pm$ 0.15 & S & $13.72$$\pm$0.01 & $<$ $11.31$ & $15.23$$\pm$0.01 & Y & 12, 17, 19\\
5 & Q2206$-$19 & 2.0762 & 20.43 $\pm$ 0.06 & $-2.25$ $\pm$ 0.07 & Si & $<$ $13.20$ & $<$ $11.83$ & $13.68$$\pm$0.03 & Y & 2, 4, 7
\enddata
\tablenotetext{a}{Ion used for metallicity determination}
\tablenotetext{b}{Part of the \citet{Neeleman2013} sample}
\tablenotetext{c}{VPFIT used to determine the column density}
\tablecomments{(This table is available in its entirety at the end of this manuscript.)}
\tablerefs{(1) \citet{Lu1996};(2) \citet{Prochaska1997};(3) \citet{Lu1998};(4) \citet{Prochaska1999};(5) \citet{Petitjean2000};(6) \citet{Prochaska2000};(7) \citet{Prochaska2001};(8) \citet{Levshakov2002};(9) \citet{Prochaska2002a};(10) \citet{Prochaska2003a};(11) \citet{Dessauges2004};(12) \citet{Khare2004};(13) \citet{Ledoux2006};(14) \citet{OMeara2006};(15) \citet{Herbert-Fort2006};(16) \citet{Dessauges2007};(17) \citet{Prochaska2007};(18) \citet{Wolfe2008};(19) \citet{Jorgenson2010};(20) \citet{Kaplan2010};(21) \citet{Rafelski2012};(22) \citet{Kulkarni2012};(23) \citet{Berg2013};(24) This Work;(25) \citet{Berg2014}}
\end{deluxetable*}

Observationally, the validity of this model has been tested for gas in the local universe. The observational studies range over a large part of the electromagnetic spectrum from X-ray \citep{Snowden1997} to radio \citep{Heiles2003a}. The results suggest that indeed some of the gas has properties of both the CNM and WNM. However, a large fraction of the WNM is actually found to be in the temperature region between 500 K and 5000 K \citep{Heiles2003b,Roy2013a, Roy2013b}. One possible explanation for the existence of this gas in what is known as the `forbidden region' comes from numerical simulations, which show that turbulence could produce the observed gas characteristics while still locally satisfying thermodynamic equilibrium \citep{Gazol2005, Walch2011}. 

To understand the evolution of galaxies, it would be ideal to measure the properties of the gas over a range of redshifts and physical conditions. This, however, is difficult to do because the methods used at low redshift are not feasible for distant galaxies. In particular, 21 cm line emission has only been detected in galaxies up to z $\sim$ 0.26 \citep{Catinella2008}. To circumvent this problem, we can study the gas in absorption against background sources (e.g. quasars). The absorbers with the largest {\hi} gas column densities are known as damped Lyman-$\alpha$ systems \citep[DLAs; for a review see][]{Wolfe2005a}. DLAs have neutral hydrogen column densities ({\nh}) equal or greater than $2 \times 10^{20}$ cm$^{-2}$, and are likely associated with galaxies as is suggested by both observations \citep[e.g][]{Wolfe2005a} and numerical simulations \citep[e.g.][]{Fumagalli2011,Cen2012,Bird2014a}.

Observational studies of DLAs have focussed mainly on line-of-sight column density measurements. Although such studies are able to measure quantities such as the metallicity \citep[e.g.][]{Rafelski2012} and the velocity structure of the absorber \citep[e.g][]{Neeleman2013}, these studies are unable to provide detailed information on the physical conditions of this gas such as the temperature and neutral hydrogen density. Several innovative methods have been devised to measure exactly these parameters for high redshift absorbers. The first method is to measure 21 cm line absorption in DLAs in front of radio-loud quasars. The integrated optical depth of the 21 cm absorption and the measured {\hi} column density will yield the spin temperature of the associated gas \citep[see][for a detailed description of this method and results]{Kanekar2014}. A second method is to measure the fine structure lines of neutral carbon, whose ratio is dependent on the physical conditions of the gas \citep{Srianand2005,Jorgenson2010}.

On the theoretical side, \citet{Wolfe2003a} extended the work of \citet{Wolfire1995} to the physical conditions pertinent to DLAs. Under the same assumptions as before, the gas in DLAs forms a two phase-medium, albeit at somewhat different density and temperatures. Observational measurements of high redshift DLAs show that indeed some of the gas has properties similar to both the CNM \citep{Howk2005,Srianand2005,Carswell2010, Jorgenson2010} and WNM \citep{Lehner2008,Carswell2012,Kanekar2014,Cooke2014}. However, disagreement lies with the percentage of DLAs that contain a significant fraction of CNM. Based on several observational results, \citet{Wolfe2003b,Wolfe2004} claim that the star formation rate per unit area is too large for current observational constraints if DLAs occur solely in a WNM \citep[see also][]{Fumagalli2014}. On the other hand the 21 cm absorption studies suggest that at least 90 \% of DLAs contain a large fraction of WNM \citep{Kanekar2003, Kanekar2014}.

To address this issue and shed additional light on the physical conditions of gas probed by DLAs, we apply in this paper a third method. This method was first described by \citet{Howk2005} for DLAs. It relies on the fact that the ratios of the fine-structure levels of the ground states of C$^+$ and Si$^+$ are solely determined by the physical parameters of the DLA \citep[see also][]{Silva2002}. Therefore a measurement of these ratios allows for a determination of the physical parameters of the DLA. This method has several advantages. Unlike \ion{C}{1}, both {\siii} and {\cii} are the dominant ionization states of these elements, and therefore they very likely trace the bulk of the neutral gas. Furthermore, unlike the 21 cm method, this method does not use a radio source, which could probe different gas, as the radio source need not be as compact as the ultraviolet or optical source \citep{Wolfe2003b, Kanekar2014}.

This paper is organized as follows. In Section \ref{sec:sample}, the selection of the sample used in this paper is explained. In Section \ref{sec:meas} we describe the measurements from the observations and literature sample. Section \ref{sec:method} explains in detail the method used in this paper to measure the physical parameters of the DLA. The results are tabulated and described in Section \ref{sec:results}. Finally we discuss these results in Section \ref{sec:disc} and summarize them in Section \ref{sec:summ}.

\section{Sample Selection}
\label{sec:sample}
To apply the method described in this paper, we require accurate measurements of the column density of the two fine structure levels of the ground state of both C$^+$ and Si$^+$. We will denote the upper level of the ground state by an asterisk (e.g. {\siiistar}), whereas the lower level will be represented by the standard notation (e.g. {\siii}). To limit saturation issues and to enable individual component analysis, we restrict ourselves to high resolution data. In particular, we limit ourselves to data from the high resolution spectrograph \citep[HIRES;][]{Vogt1994} on the Keck I telescope, which resulted in spectra with a typical resolution of $\sim$ 8 km s$^{-1}$. We further require that at least one of the transitions of both levels of the C$^+$ and Si$^+$ are clear of any forest lines or interloping features. In practice this means selecting those spectra which have clear spectral regions around {\lambdaciistar}, {\lambdasiiistar}, and {\lambdasiii}. In rare cases we use {\lambdasiiistartwo} and other {\siii} lines to determine the column densities of the Si$^+$ fine structure lines. We do not directly measure the \ion{C}{2} lower state because the \ion{C}{2} $\lambda$1334.5 line is too saturated to get accurate column densities (see Section \ref{sec:column}).

\begin{figure}[b]
\epsscale{1.2}
\plotone{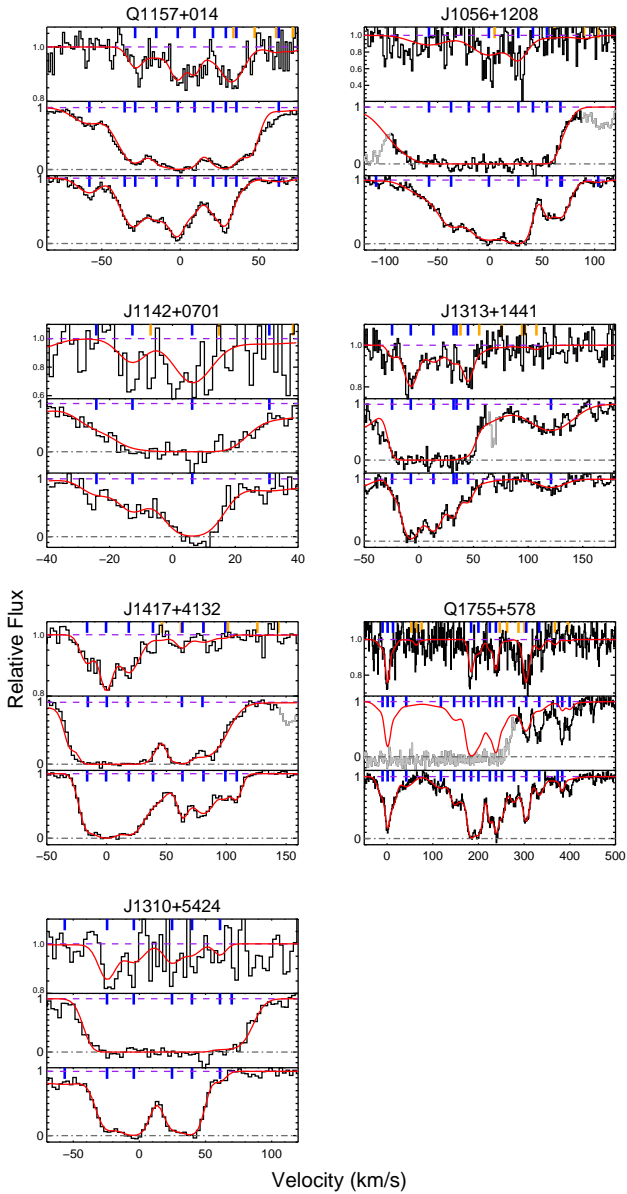}
\caption{(top) {\siiistar} , (middle) {\ciistar} and (bottom) {\siii} transitions for the 7 DLAs with detections of {\siiistar} in our sample. The dark gray (blue) vertical lines indicate the central position of the velocity components used for the line fit from $\tt{VPFIT}$ shown by the solid (red) line. In the top panel, the light gray (orange) vertical lines indicate the position of the weaker {\lambdasiiistarthree} line. Note that in all 7 DLAs the {\ciistar} transition is strongly saturated (or blended in the case of Q1755$+$578). We therefore take these fits as lower limits (see text). For the analysis discussed in this paper, we only consider the sum of the individual components, except for Section \ref{sec:ind_results} in which we discuss the component-by-component analysis.}
\label{fig:siiistar}
\end{figure} 

There are 79 spectra that satisfy the above requirements. These DLAs can in general be divided into two categories, those that where selected based on the strength of the metal lines in their Sloan Digital Sky Survey \citep[SDSS;][]{Abazajian2009} spectra \citep{Herbert-Fort2006}, and those selected purely by their {\hi} column density. We assume, as in \citet{Neeleman2013}, that the latter subset is a less biased sample and represents an accurate subsample of the general DLA population. The metal line selected sample, on the other hand, is likely biased towards higher metallicity systems and may trace the more massive host galaxy halos \citep{Neeleman2013}. In Table \ref{tab:meas} we have marked the 48 DLA that are part of the `unbiased' sample described in \citet{Neeleman2013}.

\section{Measurements}
\label{sec:meas}
This section describes the measurements taken for each absorber. The measurements for all of the DLAs in the sample are tabulated in Table \ref{tab:meas}. 

\subsection{{\hi} Column Density Measurements}
\label{sec:hi}
The {\hi} column density of the DLAs is measured by adopting the procedure outlined in \citet{Prochaska2003a}. We determine the {\hi} column density, {\nh}, of an absorber in a quasar spectrum by simultaneously fitting the continuum of the background quasar and fitting a Voigt profile to the {\lya} line of the absorber. This method provides accurate column density measurements if the continuum can be accurately placed. The measurements and their uncertainties are displayed in Table \ref{tab:meas}.

\subsection{Metallicity Measurements}
\label{sec:metal}
For each of the DLAs found, we have measured the metallicity, defined by:
\begin{equation}
[\rm{M/H}]=\log_{10}(N_{\rm{M}}/N_H)-\log_{10}(N_{\rm{M}}/N_H)_{\odot}
\end{equation}
The column density of the metals was found using the apparent optical depth method \citep[AODM;][]{Savage1991}, where we have used the wavelengths and oscillator strengths from \citet{Morton2003} and the solar abundances from \citet{Asplund2009}. We apply the same procedure as outlined in \citet{Rafelski2012}, to determine which metal to use as a tracer of the metals in a DLA. In particular, we avoid using Fe as a metal tracer for the DLAs chosen by their metal lines, as Fe is more depleted at higher metallicity \citep{Prochaska2002a, Ledoux2003, Vladilo2004, Rafelski2012}. In Table \ref{tab:meas} we list all of the metallicities of the DLAs and the line used for the determination of the metallicity.

\begin{figure*}
\epsscale{1.17}
\plottwo{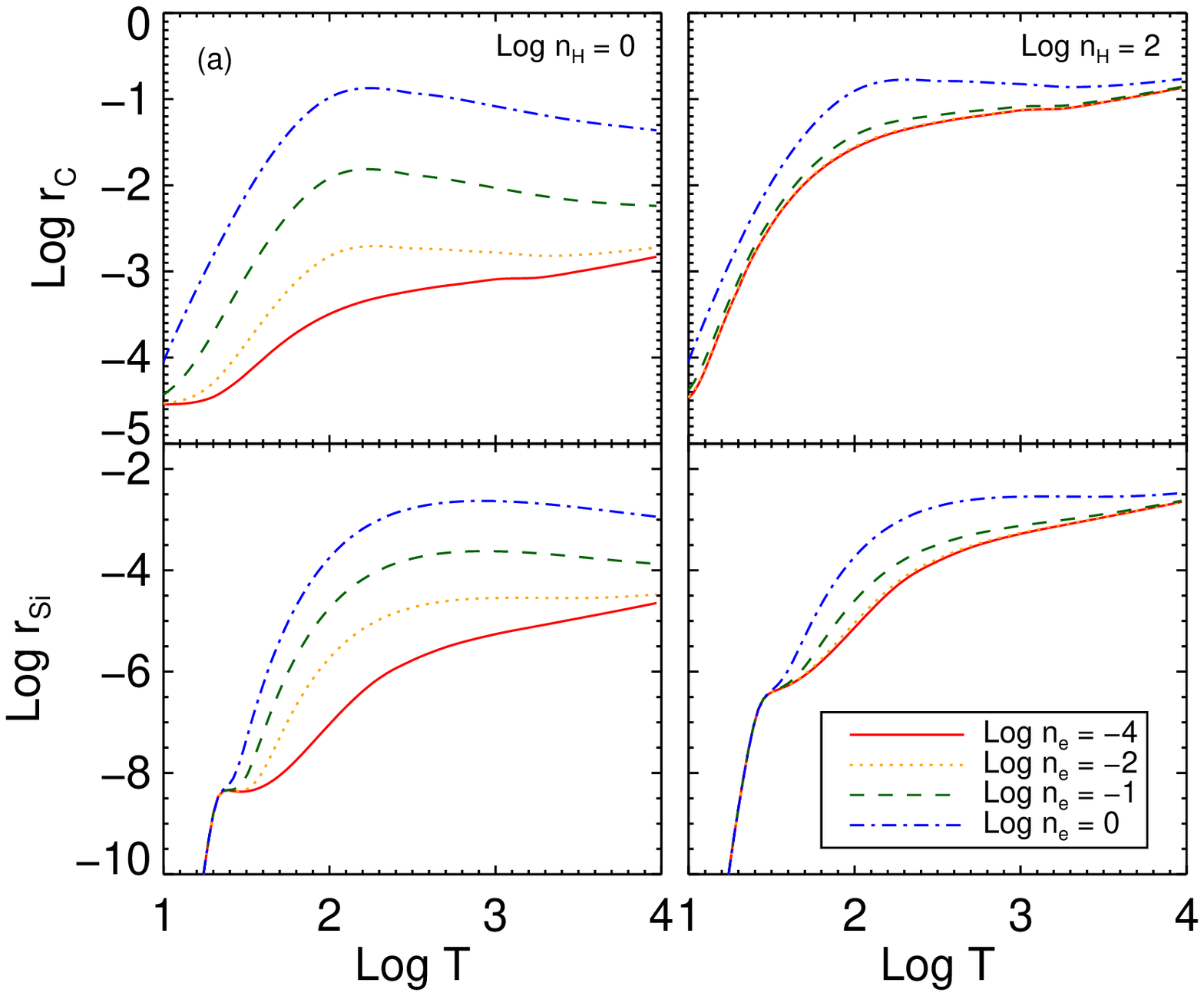}{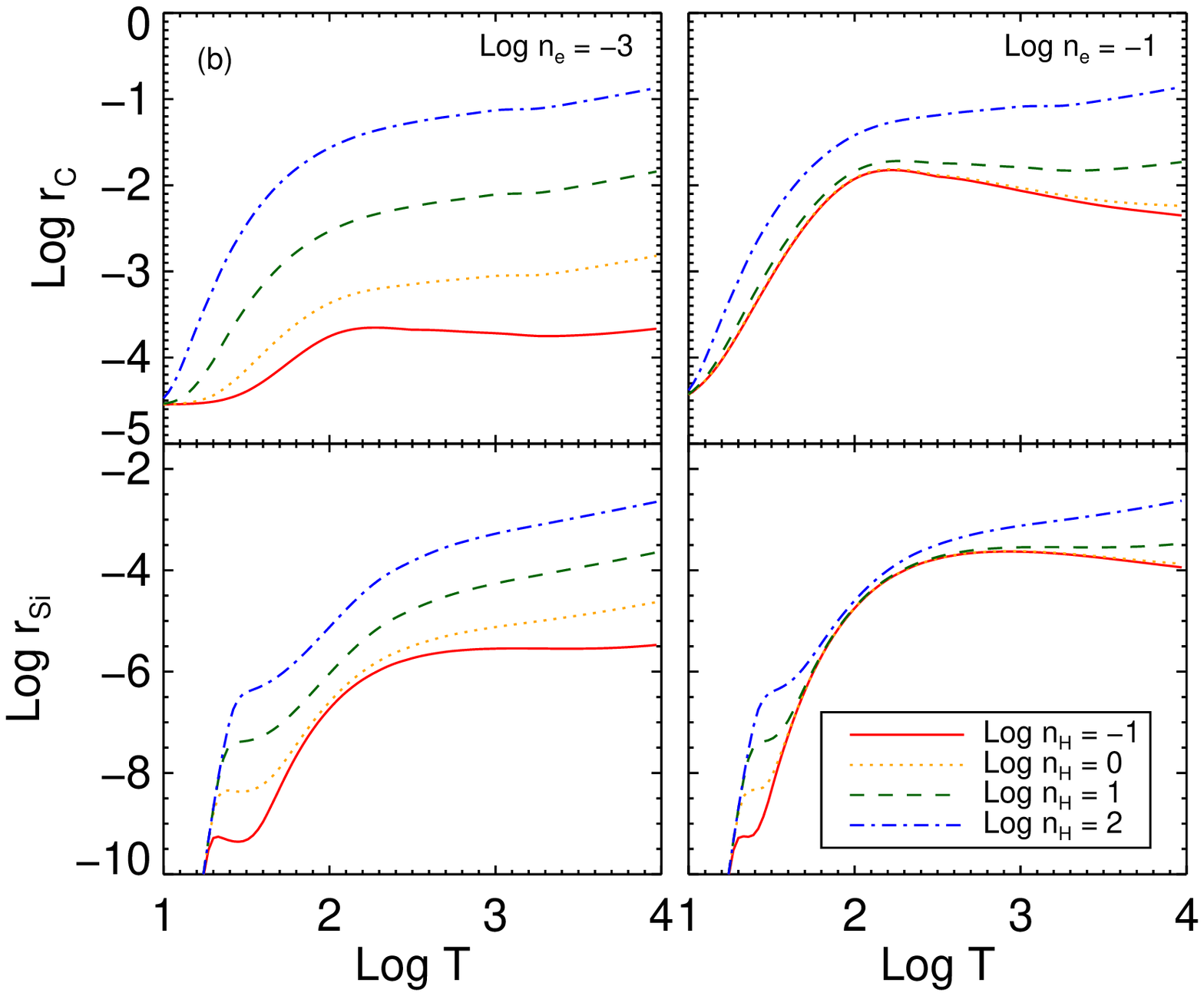}
\caption{Si and C ratios for a range of different physical parameters. Panel (a) shows how both {\rc}$=\frac{n_{\rm{C\,II^*}}}{n_{\rm{C\,II}}}$ and {\rsi}$=\frac{n_{\rm{Si\,II^*}}}{n_{\rm{Si\,II}}}$ vary with temperatures under a variety of different electron densities as the neutral hydrogen density is held constant. In panel (b) the electron density is held constant while the neutral hydrogen density is varied. These plots show that at low neutral hydrogen density, the fine-structure level ratios of C$^+$ and Si$^+$ are strongly dependent on the electron density, whereas at higher neutral hydrogen densities the ratios only dependent on the neutral hydrogen density.}
\label{fig:sicmod}
\end{figure*} 

\subsection{Column Density Measurements}
\label{sec:column}
For unsaturated lines and slightly saturated lines, the AODM provides an accurate way of determining the column density of the metals in an absorber. However, when a line is slightly blended or strongly saturated, the resultant limits for the column density found using the AODM are very conservative. Stronger constraints to the column density for such metal lines can be found by fitting the lines with $\chi^2$ Voigt profile fitting routines such as $\tt{VPFIT}$\footnote{\url{http://www.ast.cam.ac.uk/~rfc/vpfit.html}}. Since in all cases {\siiistar} is unsaturated; we only use $\tt{VPFIT}$ for our measurements of saturated {\ciistar} lines, and for {\siii} in 8 systems where all the available {\siii} lines could potentially be saturated. 

To fit the saturated lines, we select an unsaturated low-ion line such as {\siii} or \ion{Zn}{2} to determine the redshifts of the components of the low-ion lines. We then tie the Doppler parameters of the {\ciistar} and low-ion lines by assuming that they arise solely from thermal broadening. This assumption is unphysical as turbulent broadening is likely important for this gas. However, we are after a conservative lower limit to the column density \emph{independent} of the physical model used. Since the low-ion used for selecting components is heavier than the fitted ion, a thermally linked gas will provide a conservative lower limit to the column density. We finally require that the relative number of components is equal across the species and allow only the total column density of the saturated line to change. The lower limits measured using this method are marked in Table {\ref{tab:meas}.

The above methods are used to determine the column densities of {\siii}, {\siiistar}, and {\ciistar}. However, we cannot directly measure the {\cii} column density because in all cases the resonance lines of {\cii} are too saturated. We instead use the column density of Si as a proxy for C. Here we assume that both Si and C are not depleted onto dust grains and that C likely traces Fe \citep{Wolfe2003a}. Under these assumptions, [C/H]$_{\rm{gas}}$=[Fe/H]$_{\rm{int}}$=[Si/H]$_{\rm{gas}}$+[Fe/Si]$_{\rm{int}}$, where we take [Fe/Si]$_{\rm{int}}$, which is the intrinsic alpha enhancement of the DLA gas, to be $-0.3$ dex as measured by \citet{Rafelski2012}. Since both {\siii} and {\cii} are the dominant ionization state in the gas, $N$(C)$\approx$$N$({\cii}) and $N$(Si)$\approx$$N$({\siii}). Hence the column density of {\cii} is assumed to be: $\log N$({\cii})=$\log N$({\siii})$+0.62$. 

We have summarized all of these measurements in Table \ref{tab:meas}. We record a total of 7 detections of {\siiistar} from a sample of 79 DLAs. None of these detections have been analyzed previously, although two of these detections have been mentioned in the literature (i.e. J1417+4132 \citep{Berg2013}, and Q1755+578 \citep{Jorgenson2010}). The detection of 7 new {\siiistar} measurements is noteworthy, because this state is only very rarely seen in DLAs along quasar sight lines (QSO-DLAs), yet is seen regularly in DLAs detected in gamma ray bursts (GRB-DLAs). Here the GRB is likely responsible for optically pumping the excited fine structure state \citep{Prochaska2006b}. Together with the recent analysis of {\siiistar} in QSO-DLA  J1135$-$0010 by \citet{Kulkarni2012}, this sample contains all of the known detections to date of {\siiistar} in QSO-DLAs. For completeness, we have therefore included J1135$-$0010 in our sample. The 7 new {\siiistar} measurements are shown in Figure \ref{fig:siiistar}. We have also plotted a representative low-ion line to show the similarity in velocity structure between the {\siiistar} line and the low-ion lines, which is discussed further in Section \ref{sec:ind_results}.

\section{Method}
\label{sec:method}
In this section we detail the method used in this paper to determine the physical parameters of the gas using the {\siiistar} and {\ciistar} fine-structure lines. We further describe how this method is applied to all of the DLAs in our sample. 

\subsection{{\siiistar} and {\ciistar} Technique}
\label{sec:tech}
The technique of using {\siiistar} and {\ciistar} to determine the gas temperature of DLAs was first used by \citet{Howk2005}. Here we will describe the technique. The $^2P$ fine-structure states of {\siii} and {\cii} can be well approximated by a two-level atom for temperatures below 30,000 K as electrons are unable to excite the atom through collisions to higher energy levels \citep{Silva2002}. As such we can write the steady state equation for both {\siii} and {\cii} as:
\begin{equation}
\frac{n_2}{n_1}=\frac{B_{12}u_{\nu12}(z)+\Gamma_{12}+\Sigma_k n_k\gamma^k_{12}}{A_{21}+B_{21}*u_{\rm{CMB}}(z)+\Gamma_{21}+\Sigma_k n_k\gamma^k_{21}}
\label{eq:ss}
\end{equation}
Here $n_2$ refers to the $^2P_{3/2}$ state of {\siii} and {\cii} (i.e. n({\siiistar}) and n({\ciistar})), and $n_1$ refers to the lower level $^2P_{1/2}$ state of these atoms (i.e. n({\siii}) and n({\cii)). $A_{12}$, $B_{12}$ and $B_{21}$ are the Einstein coefficients for the given transitions, $u_{\rm{CMB}}$ is the energy density of the cosmic microwave background radiation field and $\Gamma_{12}$ and $\Gamma_{21}$ are the fluorescence rates. We assume the fluorescence rates are negligible because of the opacity of the ground-state transitions \citep[see e.g.][]{Sarazin1979, Wolfe2003b}, and because of the lack of Fe$^+$ excited fine-structure lines in the DLAs \citep{Prochaska2006b}. 

Finally, the excitation and de-excitation terms due to collisions (i.e. the $n_k\gamma^k_{21}$ terms) are considered. These terms are proportional to the number density of the species and the collision rate with that species. In the case of DLAs, we consider collisions with electrons, protons, and atomic hydrogen. The fraction of molecular hydrogen is assumed to be small for DLAs \citep{Jorgenson2010, Ledoux2003}, such that we can ignore collisions with this species. All of the collision rates are taken from \citet{Silva2002}, and references therein. 

Considering all these processes, the ratios of the upper to lower fine-structure levels of the ground state of C$^+$ and Si$^+$ become a function of redshift ($z$), temperature (T), neutral hydrogen density ({\nhi}), and electron density ({\nel}). Since the redshift of the DLA is well-determined from the metal lines; this leaves the three internal parameters that set the two ratios. Figure \ref{fig:sicmod} shows the dependence of the carbon ratio ({\rc}$=\frac{n_{\rm{C\,II^*}}}{n_{\rm{C\,II}}}$) and silicon ratio ({\rsi}$=\frac{n_{\rm{Si\,II^*}}}{n_{\rm{Si\,II}}}$) on temperature for a variety of different hydrogen and electron densities. The ratios correlates strongly with temperature for temperatures below 500 K. Furthermore, at low hydrogen densities ({\nhi $\lesssim$ 10 cm$^{-3}$}) collisions with electrons dominate. Hence the ratios show a strong correlation with electron density. On the other hand, at large hydrogen densities ({\nhi $\gtrsim$ 10 cm$^{-3}$})  the collisions with neutral hydrogen dominate and the ratios show a strong correlation with the neutral hydrogen density. 

Finally, we note that the observable that we measure is the ratio of the column densities, not the ratio of the actual densities. However, the two quantities are related by:
\begin{equation}
\frac{N_2}{N_1} \approx \frac{\int n_2~d\rm{s}}{\int n_1~d \rm{s}}=\frac{\int \frac{n_2}{n_1}~n_1~d\rm{s}}{\int n_1~d \rm{s}}
\end{equation}
The ratio of the two column densities is therefore simply the metal density weighted average of the density ratio over the path length. Since $d\rm{N_1} \approx n_1 d\rm{s}$, the ratio of column densities is also approximately equal to the mean of the density ratio weighted by the metal column density of the individual components. Under the assumption that the amount of neutral gas in the individual components is correlated to the column density of the components (i.e. the metallicity of the individual components is similar), then the measured ratio of the column densities will be equal to the mean of the density ratio weighted by the amount of neutral gas in each component.

We assume for this paper that the observed ratio of the column densities provides a good estimate of the density ratio for the bulk of the gas, as the column density ratio is weighted by the amount of neutral gas. One scenario where this assumption might lead to inaccurate predictions of the physical conditions of the gas is the case where the column density of the upper level arises from one phase whereas the bulk of the gas is in another phase. This is indeed expected to happen in the two-phase model where the CNM will produce the majority of {\ciistar}, even though {\cii} could come from either phase. However, requiring that the two phases are in pressure equilibrium implies the carbon density ratio, {\rc}, between the CNM and WNM differ at most by a factor of 25. Therefore even in a 60 \% WNM and 40 \% CNM mixture, the density obtained using this technique will overestimate the density for the bulk of the gas (i.e. the WNM component) by less than an order of magnitude. We explore this assumption further and compare the results from individual velocity components to each other and the system as a whole in Section \ref{sec:ind_results}.

\begin{figure}[t]
\epsscale{1.2}
\plotone{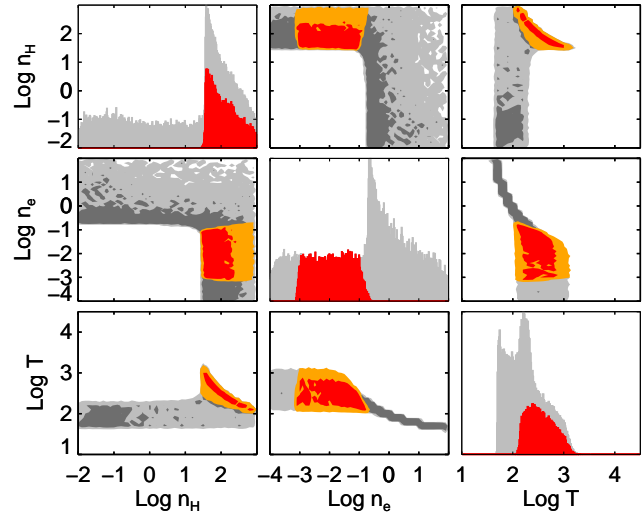}
\caption{Example of a Monte Carlo Markov Chain run for DLA J1313+1441. The upper left panel, middle panel and bottom right panel display the PDF for the individual parameters. The other panels show the parameter space that is covered by the MCMC. The light gray (dark gray) shaded regions are the 3-$\sigma$ (1-$\sigma$) boundaries for the complete chain. The gray (orange) and black (red) region are the 3-$\sigma$ (1-$\sigma$) boundaries for just the solutions that satisfy the electron density constraint described in Section \ref{sec:elec_constraint}.}
\label{fig:mcmc}
\end{figure} 

\begin{figure*}
\epsscale{1.0}
\plotone{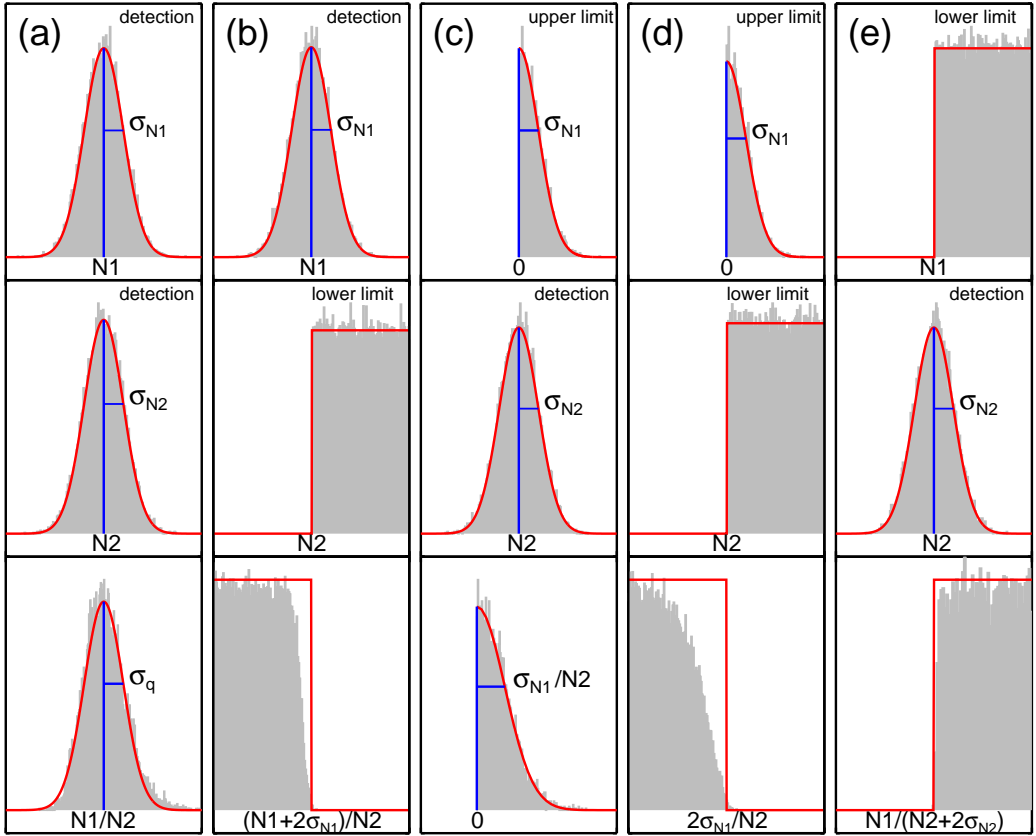}
\caption{The likelihood functions for the different possible ratios. The solid red line is the likelihood function used to describe the ratio. In gray a sample PDF with the given parameters is shown, scaled to the likelihood function. The top panel in each of the subfigures depicts the numerator, the middle panel the denominator and the bottom panel the ratio of the two. In (a) the likelihood can be approximated by a Gaussian with mean = $\frac{N1}{N2}$ and $\sigma_q=\sqrt{\left( \frac{\sigma_{N1}}{N2}\right) ^2 +\left( \frac{N1\sigma_{N2}}{N2^2}\right) ^2}$. In (c) the likelihood function of the ratio is also Gaussian. For the remaining cases (b), (c), and (e) the likelihood functions are approximated by step functions. These step functions contain $>$ 99.9 \% of the PDF. Here we have assumed a PDF for the lower limits that is flat in $\log(N)$.}
\label{fig:ratio}
\end{figure*} 

\subsection{Applying the Technique}
\label{sec:app_tech}
To explore the parameter space of all possible electron densities, neutral hydrogen densities and temperature, we apply a Monte Carlo Markov Chain (MCMC) method using the Hastings-Metropolis algorithm. This allows us to sample the complete parameter space and find the probability distribution function (PDF) for each of the physical parameters. The likelihood function used in the algorithm is discussed in Section \ref{sec:ratios}. At each step in the MCMC, this likelihood is then evaluated and multiplied by the priors. The Hastings-Metropolis algorithm is finally used to accept the step or discard it. To test for convergence, we run the MCMC five different times with varying starting points. We run the chains for $10^6$ steps, and discard the first 30 \% of the chain as our burn-in period. After the run, the PDF of each of the parameters is compared for the five different chains to check for convergence. Figure \ref{fig:mcmc} shows an example of the results for one such run.

\subsubsection{Likelihood Function and Priors of the MCMC}
\label{sec:ratios}
The likelihood function used in the MCMC method is the product of the likelihood functions of the two individual ratios:
\begin{equation}
\mathcal{L}=\prod\limits_k^{Si,C}\mathcal{L}_{\rm{k}}
\end{equation}
Here $\mathcal{L}_{\rm{k}}$ can take on different forms depending on if the column densities measured in the ratios are detections, upper limits, lower limits or some combination of the two. In our sample we have 5 different cases which are schematically shown in Figure \ref{fig:ratio}. Note that for all cases $r_{\rm{k}}$ are linear quantities, not logarithmic.

When both the numerator and denominator in the ratio are detections (Fig. \ref{fig:ratio}a), we can approximate the PDFs of the individual measurements by Gaussians. The resultant PDF of the ratio will then also be approximately Gaussian, i.e.
\begin{equation}
\label{eq:gausslike}
\mathcal{L}_{\rm{k}}=e^{-\chi_{\rm{k}}^2/2},~\rm{where}~\chi_{\rm{k}}^2=\left(\frac{r_{\rm{k,obs}}-r_{\rm{k,mod}}}{\sigma_{\rm{r_k}}}\right)^2
\end{equation}
Here $r_{\rm{k}}$ are the quotient of the upper to lower level fine structure states of C$^+$ and Si$^+$; the subscripts refer to either the observed or measured values and those from the model. $\sigma_{\rm{r_k}}$ is the uncertainty on the observed ratios calculated by standard error propagation. The prior PDF in this case is uniform. 

\begin{figure*}
\epsscale{1.15}
\plotone{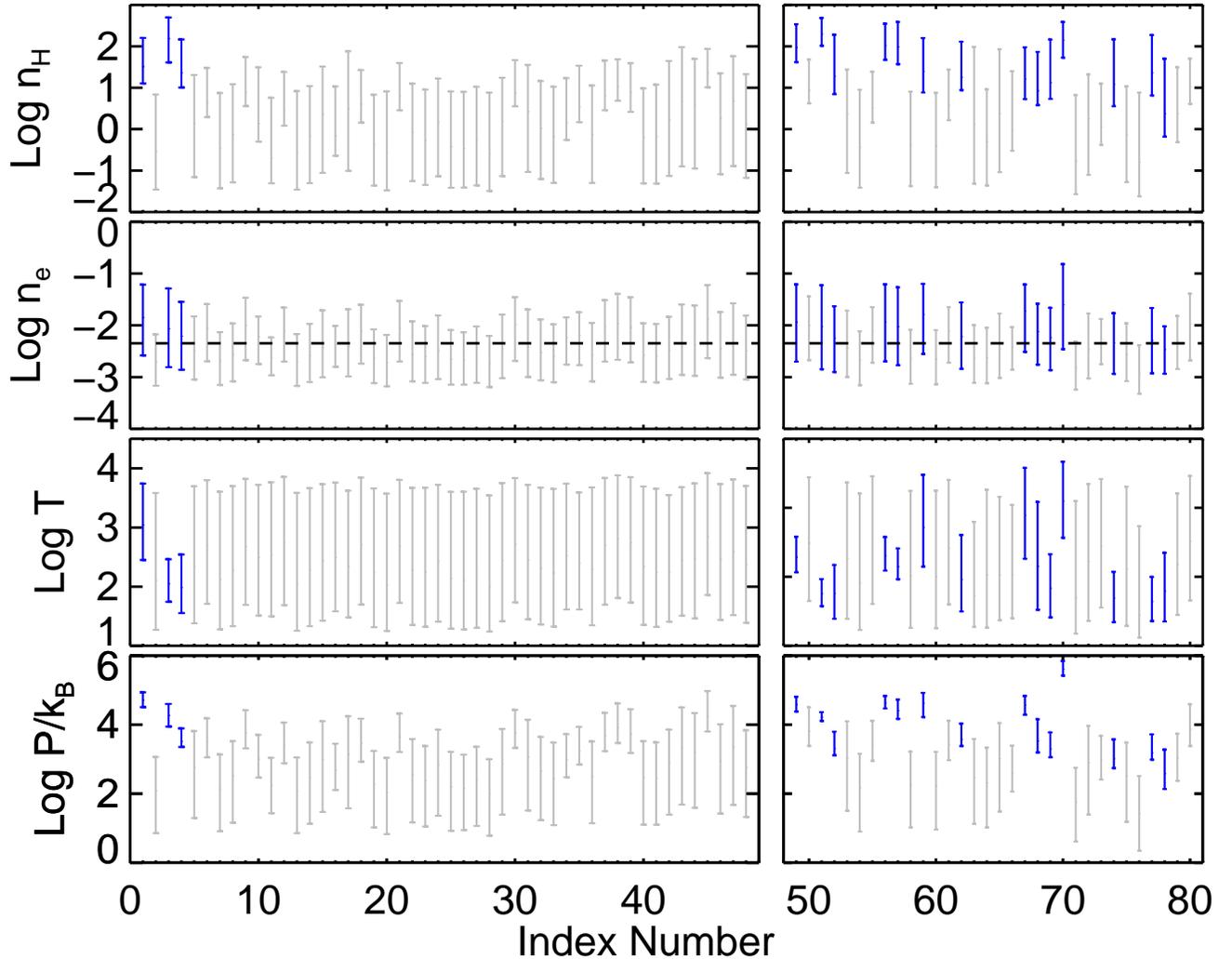}
\caption{1-$\sigma$ constraints on the neutral hydrogen density, electron density, temperature, and pressure for all of the DLAs in the sample from the {\siiistar} and {\ciistar} technique. The left panels are the results for the DLAs which are part of the unbiased sample of \citet{Neeleman2013}, whereas the right panel shows the result for the metal selected sample. The index number for the DLAs are given in Tables \ref{tab:meas} and \ref{tab:res}. The DLAs marked in black (blue) are those DLAs for which this method provides a well determined constraint on both {\nhi} and T.}
\label{fig:results}
\end{figure*} 

For upper limits on the column density measurement, we assume a Gaussian PDF centered around zero where the uncertainty is given by the 1-$\sigma$ upper limit measurement. The likelihood of a ratio consisting of an upper limit and a detection (Fig. \ref{fig:ratio}c) will then also be given by Equation \ref{eq:gausslike}. Of course negative ratios are unphysical, and therefore we assume a prior which is zero for ratios smaller than zero and uniform for ratios greater than zero.

\begin{deluxetable*}{llcccccccl}
\tabletypesize{\scriptsize}
%\rotate
\tablecaption{{Results of {\siiistar} and {\ciistar} Technique}
\label{tab:res}}
\tablewidth{0pt}
\tablehead{
\colhead{Index} &
\colhead{QSO} &
\colhead{$z_{\rm{abs}}$} &
\colhead{$\log$ {\nhi}(2-$\sigma$)} &
\colhead{$\log$ {\nel}(2-$\sigma$)} &
\colhead{$\log$ T(2-$\sigma$)} &
\colhead{$\log$ P/k$_{\rm{B}}$(2-$\sigma$)} \\
\colhead{Number} &
\colhead{} &
\colhead{} &
\colhead{[cm$^{-3}$]} &
\colhead{[cm$^{-3}$]} &
\colhead{[K]} &
\colhead{[K cm$^{-3}$]} \\
}
\startdata
1 & Q1157$+$014    & 1.9437 & $1.1$-$2.2(0.9$-$2.8)$ & $-2.6$-$-1.2(-2.9$-$-1.0)$ & $2.5$-$3.7(2.2$-$4.1)$ & $ 3.8$-$5.5( 3.5$-$6.3)$ & \\
3 & Q0458$-$02     & 2.0395 & $1.6$-$2.7(1.3$-$2.9)$ & $-2.8$-$-1.3(-3.1$-$-0.9)$ & $1.7$-$2.5(1.5$-$3.2)$ & $ 3.6$-$4.9( 3.2$-$5.6)$ & \\
4 & J2340$-$0053   & 2.0545 & $1.0$-$2.2(0.8$-$2.8)$ & $-2.9$-$-1.5(-3.2$-$-1.1)$ & $1.6$-$2.5(1.4$-$3.2)$ & $ 2.8$-$4.3( 2.5$-$5.2)$ & \\
49 & J1056$+$1208   & 1.6093 & $1.6$-$2.5(1.5$-$2.9)$ & $-2.7$-$-1.2(-3.0$-$-0.9)$ & $2.2$-$2.8(2.1$-$3.2)$ & $ 4.0$-$5.1( 3.9$-$5.6)$ & \\
51 & J0927$+$1543   & 1.7311 & $2.0$-$2.7(1.8$-$2.9)$ & $-2.8$-$-1.2(-3.1$-$-0.9)$ & $1.7$-$2.1(1.6$-$2.3)$ & $ 3.8$-$4.6( 3.6$-$5.0)$ & \\
52 & J0008$-$0958   & 1.7675 & $0.8$-$2.3(0.6$-$2.8)$ & $-2.9$-$-1.6(-3.2$-$-1.1)$ & $1.5$-$2.4(1.3$-$2.9)$ & $ 2.6$-$4.3( 2.3$-$5.0)$ & \\
56 & J1313$+$1441   & 1.7947 & $1.7$-$2.5(1.6$-$2.9)$ & $-2.7$-$-1.2(-3.0$-$-0.9)$ & $2.3$-$2.8(2.2$-$3.1)$ & $ 4.1$-$5.2( 4.0$-$5.6)$ & \\
57 & J1310$+$5424   & 1.8005 & $1.6$-$2.6(1.4$-$2.9)$ & $-2.8$-$-1.3(-3.1$-$-0.9)$ & $2.1$-$2.6(2.0$-$3.0)$ & $ 3.9$-$5.0( 3.6$-$5.4)$ & \\
59 & J1142$+$0701   & 1.8407 & $0.9$-$2.2(0.6$-$2.8)$ & $-2.6$-$-1.2(-3.0$-$-0.9)$ & $2.3$-$3.9(2.0$-$4.4)$ & $ 3.6$-$5.6( 3.1$-$6.4)$ & \\
62 & J1024$+$0600   & 1.8950 & $0.9$-$2.1(0.7$-$2.8)$ & $-2.8$-$-1.6(-3.2$-$-1.1)$ & $1.6$-$2.9(1.4$-$3.8)$ & $ 2.8$-$4.5( 2.4$-$5.6)$ & \\
67 & Q1755$+$578    & 1.9692 & $0.7$-$2.0(0.5$-$2.7)$ & $-2.5$-$-1.2(-2.9$-$-1.0)$ & $2.5$-$4.0(2.2$-$4.4)$ & $ 3.5$-$5.5( 3.2$-$6.3)$ & \\
68 & J1305$+$0924   & 2.0184 & $0.6$-$1.9(0.3$-$2.8)$ & $-2.8$-$-1.6(-3.1$-$-1.2)$ & $1.6$-$3.4(1.4$-$4.2)$ & $ 2.5$-$4.7( 2.1$-$5.9)$ & \\
69 & J1509$+$1113   & 2.0283 & $0.7$-$2.2(0.4$-$2.8)$ & $-2.9$-$-1.7(-3.2$-$-1.2)$ & $1.5$-$2.5(1.3$-$3.3)$ & $ 2.5$-$4.3( 2.2$-$5.3)$ & \\
70 & J1135$-$0010   & 2.2068 & $1.7$-$2.6(1.5$-$2.9)$ & $-2.5$-$-0.8(-2.8$-$-0.6)$ & $2.8$-$4.1(2.6$-$4.4)$ & $ 4.8$-$6.4( 4.5$-$6.8)$ & \\
74 & J0812$+$3208   & 2.6263 & $0.6$-$2.2(0.2$-$2.8)$ & $-2.9$-$-1.8(-3.2$-$-1.2)$ & $1.4$-$2.3(1.3$-$2.7)$ & $ 2.2$-$4.1( 1.8$-$4.9)$ & \\
77 & J2100$-$0641   & 3.0924 & $0.8$-$2.3(0.5$-$2.8)$ & $-2.9$-$-1.7(-3.2$-$-1.1)$ & $1.4$-$2.2(1.3$-$2.5)$ & $ 2.5$-$4.1( 2.1$-$4.8)$ & \\
78 & J1155$+$0530   & 3.3260 & $-0.2$-$1.7(-0.7$-$2.7)$ & $-2.9$-$-2.0(-3.2$-$-1.3)$ & $1.4$-$2.6(1.2$-$3.8)$ & $ 1.5$-$3.8( 1.0$-$5.2)$ & 
\enddata
\tablecomments{(This table is available in its entirety at the end of this manuscript.)}
\end{deluxetable*}

Lower limit measurements of the column density are more difficult to deal with, as it is hard to estimate an appropriate uncertainty on the measurement, because the uncertainty is strongly dependent on the model used to fit the absorption line. If the lower limit measurement is in the denominator of the silicon or carbon ratio (Fig. \ref{fig:ratio}b and d), the resultant measured ratio is an upper limit. In this case, ratios derived from the model should have a high likelihood when they are smaller than the measured ratio and small when they are bigger than the measured ratio. We therefore adopt the following conservative likelihood function:
\begin{equation}
\label{eq:steplike}
\mathcal{L}_{\rm{k}}=\left\{
\begin{array}{l l}
1 \quad r_{\rm{k,mod}} \leq r_{\rm{k,obs}} \\
0 \quad r_{\rm{k,mod}} > r_{\rm{k,obs}} \\
\end{array}
\right.
\end{equation}
This step function rules out models that produce ratios greater than $r_{\rm{k,obs}}$, and will give equal likelihood for all ratios below $r_{\rm{k,obs}}$. Here $r_{\rm{k,obs}}$ is the 2-$\sigma$ upper limit of the detection or upper limit divided by the lower limit. As Figure \ref{fig:ratio}b and \ref{fig:ratio}c show, this is a conservative approach, as $>$99.9\% of a mock generated PDF falls below this limit. Again the prior is assumed to be uniform for ratios greater than zero, and zero for ratios smaller than zero.

Similarly, for the few cases where the lower limit is in the numerator (Fig. \ref{fig:ratio}e) we use the following likelihood function:
\begin{equation}
\label{eq:steplike2}
\mathcal{L}_{\rm{k}}=\left\{
\begin{array}{l l}
0 \quad r_{\rm{k,mod}} < r_{\rm{k,obs}} \\
1 \quad r_{\rm{k,mod}} \geq r_{\rm{k,obs}} \\
\end{array}
\right.
\end{equation}
This step function rules out all models that produce ratios smaller than $r_{\rm{k,obs}}$. Here $r_{\rm{k,obs}}$ is the lower limit divided by the 2-$\sigma$ upper limit of the detection. Note that in these cases we assume a uniform prior.

\subsection{Electron Density Constraint}
\label{sec:elec_constraint}
When we apply the MCMC chain, we allow the temperature, hydrogen density and electron density to vary independently. This can clearly result in unphysical situations for DLAs where we expect the fractional ionization to be significantly smaller than 1. To apply this constraint, we assume that the fractional ionization of hydrogen, $x(\rm{H}^+)$, satisfies the following steady state equation \citep{Draine2011}:
\begin{equation}
\label{eq:elec}
\begin{split}
\zeta_{\rm{CR+X}}(1+\phi_s)[1-x(\rm{H^+})]=\mspace{200mu}\\
\alpha_{\rm{rr}}(\rm{H}^+)n_{\rm{H}}^2[x(\rm{H^+})+x(\rm{M^+})]x(\rm{H^+})+\\
\alpha_{\rm{gr}}(\rm{H}^+)n_{\rm{H}}^2x(\rm{H^+})\mspace{145mu}
\end{split}
\end{equation}
Here, $\zeta_{\rm{CR+X}}$ is the primary ionization rate of both cosmic rays and strong X-rays, $\phi_s$ are the secondary ionization rates, $\alpha_{\rm{rr}}$ is the rate coefficient for radiative recombination of H$^+$ which is a function of temperature, and $\alpha_{\rm{gr}}$ is the effective rate coefficient for grain-assisted recombination of H$^+$, which is a function of temperature, electron density and the UV radiation field. The ionization of metals is assumed to be the same as it is for local ISM scaled to the metallicity of the DLA. Using the estimates for these parameters in \citet{Draine2011}; and references therein, we can make an estimate for the electron density as a function of the neutral hydrogen density, temperature, primary ionization rate and the UV radiation field.

This constraint is imposed upon the Monte Carlo Markov Chain after the run. All the values that do not satisfy the above equation are rejected. Since the value of $\zeta_{\rm{CR+X}}$ at these redshifts is uncertain \citep[see e.g.][]{Dutta2014}, and the UV radiation field can likely take on a wide range of values depending on the separation between the DLA and any potential star forming region, we allow for a wide range of acceptable values. Specifically, the UV radiation field may range between 0.1 and 100 $G_0$ (Habing's constant; $G_0 =1.6 \times 10^{-3} \rm{ergs~cm}^{-3} \rm{~s}^{-1}$) and $\zeta_{\rm{CR+X}}$ between $10^{-17}$ and $10^{-15}$ $s^{-1}$. 

As a result of this constraint, the electron densities never exceed densities of about 0.1 cm$^{-3}$, because such electron densities would require the gas to be significantly ionized. Similarly, electron densities below $\sim 10^{-4}$ cm$^{-3}$ are ruled out because of the intrinsic electron density due to the singly-ionized metals such as carbon. One example of the application of this constraint to the MCMC chain are shown by the orange and red contours in Figure \ref{fig:mcmc}.

\section{Results}
\label{sec:results}
This section describes the results from the {\siiistar} and {\ciistar} technique. The output of the technique is a PDF on each of the three physical parameters (e.g. {\nhi}, {\nel} and T). We have plotted the 1-$\sigma$ ranges for each of these parameters in Figure \ref{fig:results}; we also have included the pressure constraints of these systems in this figure (see Section \ref{sec:param}). The results are tabulated in Table \ref{tab:res}. Not all DLAs have well-determined ranges on all internal parameters, either due to low S/N spectra or because the resultant ratios are not strongly correlated with a specific internal parameter (see Section \ref{sec:tech}). Those DLAs that have well-determined ranges in both neutral hydrogen density and temperature are plotted in blue in Figure \ref{fig:results} and are shown in the abbreviated version of Table \ref{tab:res}. 

\subsection{Physical Parameters of DLAs}
\label{sec:param}
The top panel of Figure \ref{fig:results} shows the distribution of neutral hydrogen density in our sample. The range of allowed neutral hydrogen column densities varies significantly between DLAs. Several DLAs such as J0927$+$1543 are very dense, with 1-$\sigma$ lower limits on the density of 100 cm$^{-3}$. These high values are driven by high {\rc}, and the non-detection of {\siiistar}. On the other hand, several other DLAs have 1-$\sigma$ upper limits of 10 cm$^{-3}$ indicating that the gas in DLAs exhibits a wide range of neutral hydrogen density. It is important to note that the {\siiistar} and {\ciistar} method cannot precisely measure neutral hydrogen densities below 1 cm$^{-3}$, because for these densities the interactions with neutral hydrogen becomes subdominant to collisions with electrons. As such, only DLAs with neutral hydrogen densities above this value have well-determined constraints on their neutral hydrogen density.

The second panel of Figure \ref{fig:results} shows the distribution of electron densities. One interesting feature of this distribution is that the variation in the range of electron densities is significantly less compared to the range of neutral hydrogen densities. This is a direct consequence of applying Equation \ref{eq:elec}, which makes the electron density only weakly dependent on the temperature and density of the gas. Specifically, Equation \ref{eq:elec} gives an electron density of {\nel} = 0.01 cm$^{-3}$ for both a canonical CNM ({\nhi} = 30 cm$^{-3}$ and T = 50 K) and WNM ({\nhi = 0.5 cm$^{-3}$ and T = 5000 K) for local ISM conditions. Note that this is slightly higher than the median value of the complete DLA sample (n$_{\rm{e}}=0.0044 \pm 0.0028 \rm{cm}^{-3}$). The difference is likely due to the lower metallicity of the DLA sample and different UV radiation fields and ionization rates from cosmic rays and X-rays \citep{Wolfire1995}. As a result, we are unable to differentiate between the electron densities of DLAs, but a typical DLA will have an electron density of about 4 $\times 10^{-3}$ cm$^{-3}$, which is consistent with the values found by \citet{Srianand2005}.

The third panel of Figure \ref{fig:results} shows the temperature range for each of the DLAs in our sample. Several DLAs have temperature ranges that are consistent with the temperatures expected from a CNM. To be specific, nine of the DLAs have 1-$\sigma$ upper limits on the temperature of 500 K. On the other hand, there are several DLAs that have ranges that are at significantly higher temperature. We again would like to stress that the {\siiistar} and {\ciistar} method is unable to measure the precise temperature above 500 K as {\rc} and {\rsi} become weak functions of temperature. Therefore only the coldest DLAs have well determined ranges on their temperature.

Finally, in the bottom panel of Figure \ref{fig:results} we have plotted the pressure range for each of the DLAs. The pressure was calculated from the MCMC chains by taking the product of the neutral hydrogen density and the temperature, since $P/k_{\rm{B}}=n_{\rm{H}}T$. As pressure shows the strongest correlation with {\rsi} and {\rc}, it is the best constrained parameter. The median pressure for the complete sample is $\log P/k_{\rm{B}} =3.0 $ [K cm$^{-3}$]. This is lower compared to the pressure of the ISM measured locally using the \ion{C}{1} method \citep{Jenkins2011}. We discuss this further in Section \ref{sec:disc_compdat}.

In Figure \ref{fig:tvsnh} we have plotted the temperature versus the density for the complete sample of DLAs. We have also indicated the typical ranges for a canonical CNM, WNM and the classically forbidden region defined by the two-phase model \citep[see e.g.][]{Heiles2012}. Nine DLAs with well-determined ranges on the temperature and neutral hydrogen density have physical conditions that are consistent with those expected from gas in a CNM. Two of these DLAs are from the 48 DLAs which are part of the unbiased sample of \citet{Neeleman2013}. We therefore conclude that at least 5\% of a random sample of DLAs contain significant fractions of CNM. This percentage is a lower limit because many DLAs with less well-determined ranges are consistent with gas in a CNM as is shown by the gray and dark gray contours. 

The remaining DLAs are spread over a wide range of temperature and neutral hydrogen densities, all of them consistent with both canonical CNM and WNM conditions. In Figure \ref{fig:tvsnh} the 68 \% and 99 \%  confidence contours for the temperature and neutral hydrogen density of the remaining DLAs are shown. These contours rule out the parameter space of high neutral hydrogen density and high temperature, as such physical conditions would produce too large {\rsi} and {\rc} (see Figure \ref{fig:sicmod}).

\begin{figure}
\epsscale{1.25}
\plotone{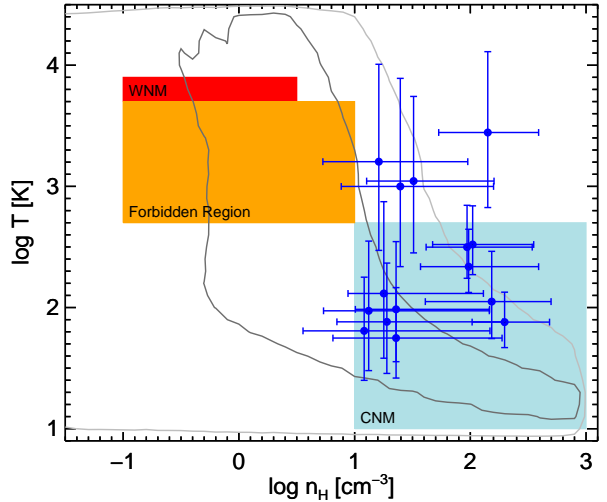}
\caption{Temperature and neutral hydrogen density for the complete sample. The data points are those DLAs with well-defined ranges on {\nhi} and T (see Figure \ref{fig:results}. The dark gray (light gray) lines indicate the 68 \% (99 \%) confidence interval of the neutral hydrogen density and temperature for the remaining DLAs. We have also plotted the typical ranges for a canonical cold neutral medium, warm neutral medium, and the classically forbidden region.}
\label{fig:tvsnh}
\end{figure} 

\begin{figure*}
\epsscale{1.2}
\plotone{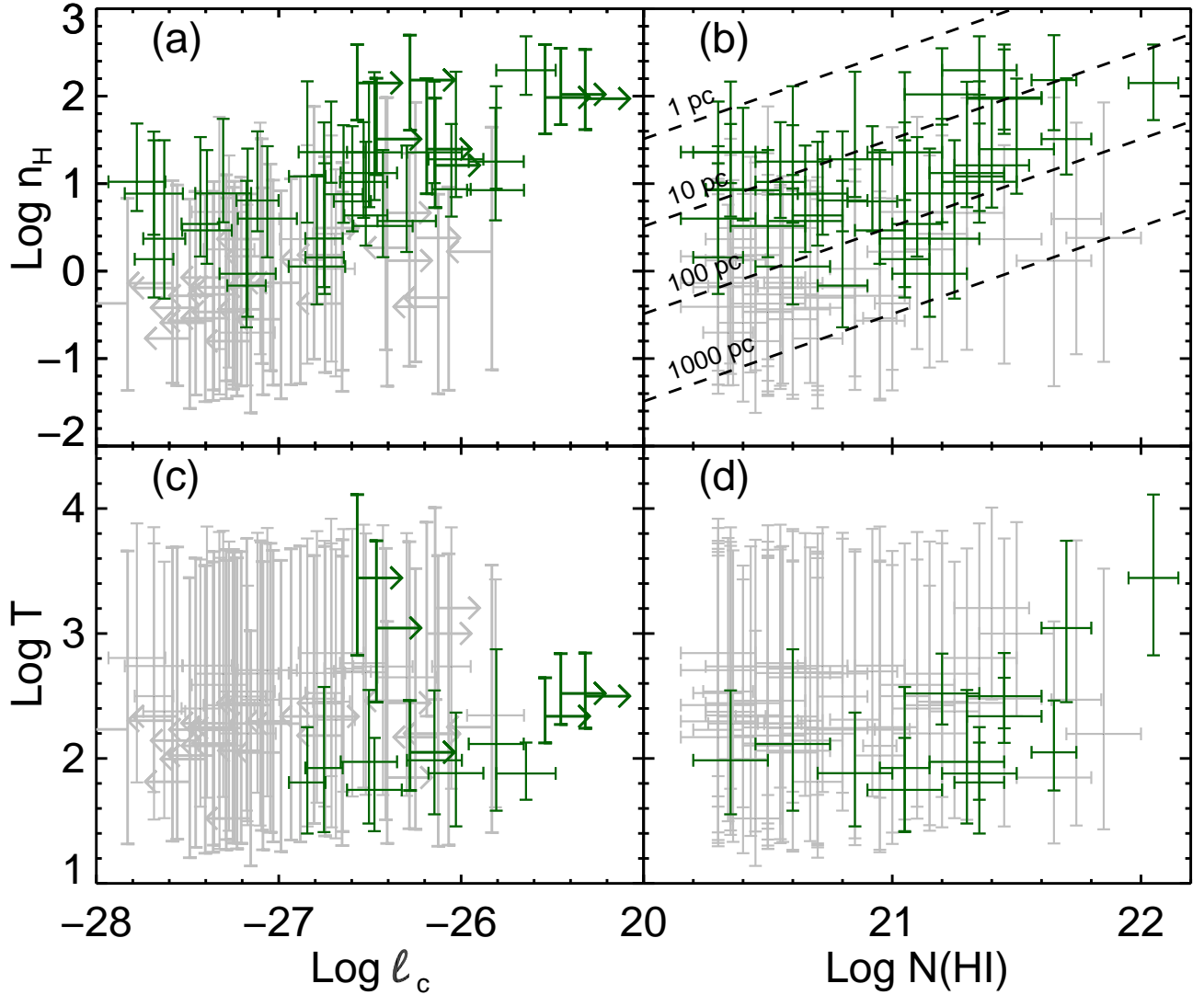}
\caption{Selection of possible correlations seen in the data. The black (green) data points are those points with strong constraints on their physical parameter from the MCMC analysis. Panel (a) shows the correlation between cooling rate and the hydrogen density. This correlation is expected as larger cooling rates indicate larger {\ciistar} ratios which result in higher neutral hydrogen densities. Panel (b) plots hydrogen density vs neutral hydrogen column density. This plot indicates a maximum cloud size of DLAs of less than 1 kpc. The last two panels show that temperature is not strongly correlated with any of the external parameters.}
\label{fig:corr}
\end{figure*} 

\begin{figure}[b]
\epsscale{1.25}
\plotone{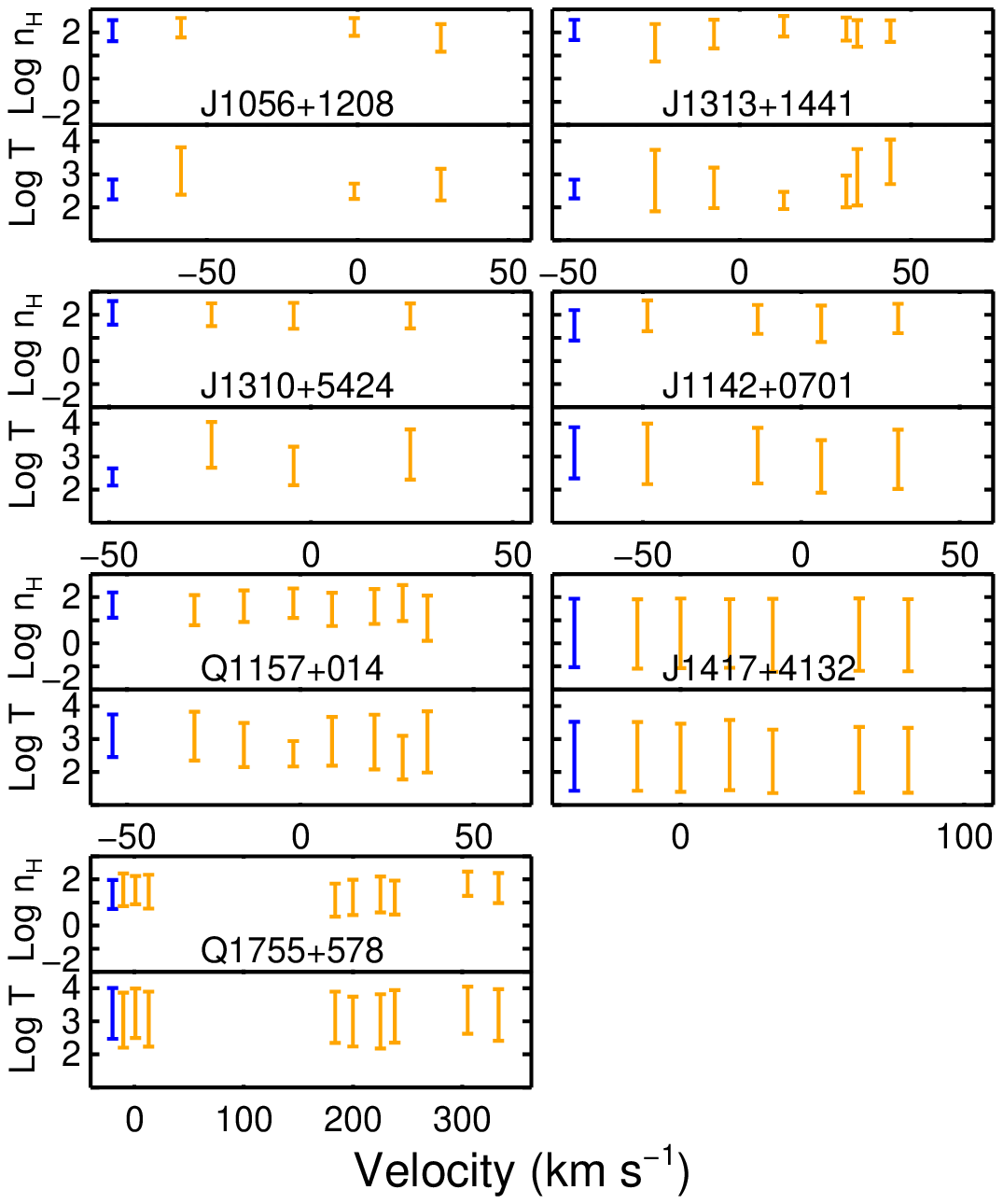}
\caption{Temperature and neutral hydrogen density for the individual components for those DLAs with detectable levels of {\siiistar}. The gray (orange) values are the measurements for the hydrogen density and temperature for the individual components, and are placed at the velocity center of the component. The black (blue) data point is the result from the MCMC analysis by considering the system as a whole.}
\label{fig:comp}
\end{figure} 

\subsection{Correlations with Global DLA Properties}
In this section we explore possible correlations between the physical parameters measured with the {\siiistar} and {\ciistar} method and the global properties of the DLAs. We consider all of the global DLA parameters discussed in \citet{Neeleman2013}. Figure \ref{fig:corr} shows a selection of these correlations. As panel (a) of this figure shows, there is a clear correlation between the neutral hydrogen density and the 158 $\mu$m cooling rate of the DLAs \citep[{\elc}; see e.g.][]{Wolfe2003a}. This result is due in part because the cooling rate is proportional to the {\ciistar} column density and larger {\ciistar} column densities result in larger {\rc}, which in turn yield higher neutral hydrogen column densities (see Figure \ref{fig:sicmod}). We consider this result evidence for the two-phase model described by \citet{Wolfe2003a}, where higher star formation rates, and therefore higher cooling rates, result in higher stable equilibrium densities for hydrogen. Since {\elc} is correlated to metallicity, redshift and the kinematical parameters \citep{Wolfe2008,Neeleman2013}, the neutral hydrogen density also shows a correlation (albeit weaker) with these parameters. 

Panel (b) is a plot of neutral hydrogen density vs {\hi} column density. The ratio of these components gives a crude estimate of the absorption length of the DLA. To be specific, the {\siiistar} and  {\ciistar} technique provides an estimate of the density of the bulk of the neutral gas (Section \ref{sec:tech}). Therefore the ratio of the neutral hydrogen density and {\hi} column density will give an upper limit to the size of the component which contains the bulk of the neutral gas. The results show that for the majority of DLAs the absorption lengths for these components is less than 1 kpc. Indeed some absorption lengths are as small as a few tens of parsec. These very small absorption lengths correspond in general to those DLAs with well determined cold temperatures (T $\le$ 500 K). This suggests that for these DLAs the bulk of the gas is located in relatively small cold components. 

Unlike the hydrogen density, the electron density shows no significant correlation with any of the global DLA parameters. Similarly, the temperature shows no significant correlation either. However, we would like to discuss two interesting features of the temperature measurements. The first feature is that the DLAs with the highest cooling rate have on average a lower temperature range (Figure \ref{fig:corr}c). This likely is related to the correlation between cooling rate and neutral hydrogen density, as higher neutral hydrogen densities likely correspond to colder environments (Figure \ref{fig:tvsnh}). The second feature is that the two highest column density systems have the highest limits on the temperature (Figure \ref{fig:corr}d). A possible explanation for this result is that low temperature gas will form molecular gas, limiting the maximum allowed column density of neutral atomic hydrogen \citep{Schaye2001b}. The molecular fraction is believed to be anti-correlated with temperature \citep{Schaye2001b, Richings2014b} and therefore larger atomic neutral hydrogen column densities are possible for higher temperature DLAs. 

\subsection{Component Analysis}
\label{sec:ind_results}
In all of the analysis we have assumed that the ratio of column densities is approximately equal to the ratio of the densities of the bulk of the neutral gas. To explore this assumption, we have repeated the analysis on each of the individual velocity components of those DLAs with measurable {\siiistar} (Figure \ref{fig:siiistar}). The results are displayed in Figure \ref{fig:comp}. As can be seen from the individual panels, the majority of the velocity components have physical parameters that are within 1-$\sigma$ equal to the measurements from treating the system as a whole. This result is due to the similarity in the column density ratio between the upper and lower levels of the fine structure states of {\siii}. Note that in Figure \ref{fig:siiistar} the {\siiistar} line traces the low-ion line quite well for the majority of DLAs. There are several exceptions such as the component at $+$50 {\kms} for DLA J1313+1441. This component has an {\rsi} five times greater than the mean value of the DLA, resulting in a temperature range inconsistent with that found for the total DLA. However, such components are uncommon; the mean deviation in {\rsi} and {\rc} from component to component is less than 50 \% of the mean value, which results in similar ranges for the physical parameters.

The similarity between {\rsi} and {\rc} for the different velocity components strengthens the assumption to take the ratio of the total column densities to be equal to the ratio of the densities, since a per component analysis will produce similar results. One possible explanation for the similarity between the individual velocity components is that the distinct components are physically close to each other and experience similar exterior physical conditions, or a second explanation could be that external conditions are similar over a large portion of the absorbing galaxy.

There are three caveats to this results. The first caveat is that this result does not exclude the existence of \emph{any} clumps of gas with strongly varying physical parameters along the quasar line of sight. It does, however, suggest that these clumps can only contribute a very small fraction of the total metal column density, and therefore are not likely to describe the bulk of the neutral gas. The second caveat is that the DLAs used in the individual component analysis all have measurable levels of {\siiistar} and therefore might not be representative of the DLAs in general. We have tested this caveat by considering the ratio of {\ciistar} to {\siii} in a sample of DLAs for which both transitions are detected, and we find that this ratio is also not strongly varying between the individual components \citep[see also][]{Wolfe2003a,Wolfe2003b}. Therefore we believe that this is a general result holding for the majority of DLAs. Finally, the third caveat is that it could be that the individual velocity components are in actuality composed of a collection of smaller components. In this case, each individual component is averaged in a similar manner as the whole DLA, and therefore weighted most strongly by the component with the largest metal column density. As a result the {\siiistar} and {\ciistar} technique will still recover the physical conditions of the bulk of the gas.

\section{discussion}
\label{sec:disc}
The discussion section is organized as follows. In Section \ref{sec:disc_data} we discuss the empirical results. We compare these results to previously measured data for both DLAs and the local ISM in Section \ref{sec:disc_compdat}. In Section \ref{sec:disc_compmod} we discuss how the results of this paper fit into models describing the ISM, in particular the two-phase model. Finally, in Section \ref{sec:disc_impl} we will comment on what these results suggest for the ISM of high redshift galaxies.

\subsection{Discussion of the Empirical Results}
\label{sec:disc_data}
We found in Section \ref{sec:results} that the temperature distribution of the full sample cannot precisely be determined because the Si$^+$ and C$^+$ ratios are insensitive to temperature changes when T exceeds 500 K. However, we can measure the minimum fraction of DLAs that have gas temperatures consistent with a CNM (i.e. T $<$ 500 K). Using the `unbiased' subsample of \citet{Neeleman2013}, we find that at least 5 \% of DLAs have the bulk of their neutral gas in cold, dense clouds with conditions similar to a CNM (Figure \ref{fig:tvsnh}). We again note that this is a strict lower limit, as DLAs with less well-determined ranges on their physical parameters would increase this percentage. 

As quasar lines of sight randomly probe the gas surrounding the DLA galaxy, we can convert this percentage to an approximate volume filling fraction of CNM in high redshift DLA galaxies. Depending on the exact geometry and distribution of the CNM, the minimum volume filling fraction must be at least 1 \%.  This is very similar to the CNM filling fraction for the local ISM which is found to be approximately 1 \% \citep{Draine2011}, and indicates that the volume filling fraction of CNM for high redshift galaxies is at least in rough agreement with the value measured for low redshift galaxies.

\begin{figure}[b]
\epsscale{1.25}
\plotone{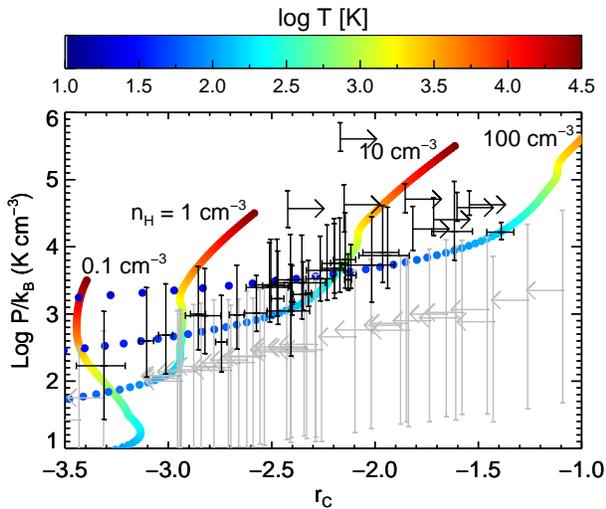}
\caption{Correlation between pressure and {\rc}. The black data points are lower limits and measurements of {\rc}, whereas the gray data points are 2-$\sigma$ lower limits. The grayscale (colored) tracks are the theoretical solutions to {\rc} and pressure for the given neutral hydrogen density and temperature assuming an electron density of {\nel} = 0.0044 cm$^{-3}$. Most {\rc} detections, which account for about 40 \% of a random DLA sample, are inconsistent with the conditions found in a canonical WNM.}
\label{fig:rcp}
\end{figure} 

In Section \ref{sec:param}, we showed that pressures vary significantly between DLAs and is correlated to {\rc}. This correlation is shown in Figure \ref{fig:rcp}. The tracks are theoretical temperature paths for the indicated neutral hydrogen density assuming the measured median electron density of 0.0044 cm$^{-3}$. The black data points are detections or lower limits to {\rc}. This figure illustrates two things. Firstly, the majority of detections and lower limits in {\rc} cannot be produced in neutral gas with {\nhi} $\lesssim$ 1 cm$^{-3}$, which means that a canonical WNM is rarely able to produce detectable levels of {\ciistar}, and because {\rc} values can be measured in approximately 40 \% of DLAs \citep[see][]{Neeleman2013} this indicates that these DLAs must contain some fraction of gas not in a canonical WNM. Secondly, the DLAs with the highest C$^+$ ratios have pressures and neutral hydrogen densities significantly higher than the median, and are in general those DLAs with well-defined limits on temperature and density.

Indeed, if we consider just the systems with well-defined ranges on temperature and density, we find that these systems show significantly higher velocity widths with a median velocity with of 131 {\kms} which is almost double the median value of a random DLA sample \citep{Neeleman2013}. They also show an increased metallicity and cooling rate, all suggesting that these systems are part of the most massive dark matter halos which give rise to DLAs (see further Section \ref{sec:disc_implgal}). 

Finally a detailed look at the individual velocity components of DLAs shows that there exists little differences between the measured ratios between the individual velocity components. As was suggested in Section \ref{sec:ind_results} this could be due to close proximity of the individual components. In particular, the observed {\rc}$=${\ciistar}/{\cii} is proportional to:  
\begin{equation}
r_{\rm{C}} \propto \frac{\ell_c}{\rm{[M/H]}}
\end{equation}
Hence, if we assume that the metallicity of the individual components does not vary significantly, then the constancy in {\rc} indicates that the cooling rates are the same across the individual components. This is indeed expected and assumed in the two phase model of \citet{Wolfe2003a}, as they found that heating rate (and because of the assumed thermal equilibrium therefore also cooling rate) is a global property of the DLA and not a local property as it is in the local ISM. The constancy of {\rc} across the different components is therefore expected in the two-phase model, as it is a direct consequence of the global nature of the heating rate.

\subsection{Comparison with Previous Observations}
\label{sec:disc_compdat}
In Table \ref{tab:res} we have listed the results of our study on the neutral hydrogen density, electron density, temperature, and pressure of DLAs using the {\siiistar} and {\ciistar} method. This is not the first study of these parameters as several other methods provide estimates. A comparison for those 7 DLAs which have measured temperatures from multiple methods is shown in Table \ref{tab:comp}. These DLAs suggest that there is a reasonable agreement between the physical parameters derived using the {\siiistar} and {\ciistar} method and previous methods.

\begin{deluxetable}{llllllll}
\tabletypesize{\scriptsize}
%\rotate
\tablecaption{{Comparison between Temperature Measurements}
\label{tab:comp}}
\tablewidth{0pt}
\tablehead{
\colhead{QSO} &
\multicolumn{3}{c}{T (K) (1$-\sigma$ constraint)} \\
\cline{2-4}
\colhead{} &
\colhead{{\ciistar}/{\siiistar}} &
\colhead{Other} &
\colhead{Method}
}
\startdata
Q0336$-$01   & (28 - 5200)    & $>$ 8890        & 21 cm absorption \\
Q0458$-$02   & (50 - 316)     & (465 - 655)    & 21 cm absorption \\
Q1157$+$014  & (260 - 5500)   & (760 - 1270)   & 21 cm absorption \\
J0812$+$3208 & (25 - 178)     & (32 - 88)      & \ion{C}{1} \\
J2100$-$0641 & (30 - 171)     & (10 - 251)     & \ion{C}{1} \\
Q2206$-$19   & $>$ 25         & (9200 - 15200) & line-fitting \\
J2340$-$0053 & (36 - 375)     & (55 - 200)     & \ion{C}{1}
\enddata
\end{deluxetable}

\subsubsection{21 cm Absorption}
As was discussed in the introduction, one method of measuring the spin temperature of DLAs is by measuring 21 cm absorption in DLAs located in front of radio-loud quasars. A comprehensive paper describing this method was recently published by \citet{Kanekar2014}. They found that the median temperature of the gas inside DLAs responsible for 21 cm absorption in their sample was greater than 900 K. Furthermore, they found that only 2 out of the 23 DLAs above a redshift of 1.7 were consistent with having a significant fraction of CNM. We note that we found in our sample that this fraction must be at least 5 \%. These results are consistent within the uncertainty of the measurements due to the small sample sizes of both methods. Because the former measurement is an upper limit and the latter a lower limit, the two methods suggest that roughly between 5 and 10 \% of all DLAs have the bulk of their gas in a CNM phase.

This fraction is somewhat in conflict with the results from \citet{Wolfe2003b,Wolfe2004}, who argued that the majority of all {\ciistar} detections in DLAs must come from gas in a CNM, and about 40 \% of all DLAs have detectable levels of {\ciistar} \citep[e.g.][]{Neeleman2013}. \citet{Kanekar2014} resolves this conflict by assuming that only a small fraction of the gas (10 - 20 \%) in the DLAs with {\ciistar} detections is in actuality CNM, with the bulk of the gas in a WNM phase. There are two problems with this scenario. First, it is unclear why in this scenario, the {\ciistar}/{\siii} ratio would be relatively constant among the individual velocity components, as it is in observations. Secondly, using the technique described in this paper, we can calculate the amount of {\cii} needed in the CNM to produce the required amount of {\ciistar} observed. In the two cases mentioned in \citet{Kanekar2014} (i.e. DLAs Q1157$+$014 and Q0458$-$02), the resultant {\cii} column density needed in a canonical (T = 100 K) CNM to produce the observed {\ciistar} column density is larger than the observed \emph{total} {\cii} column density. Hence, at least for these two DLAs, we can rule out a scenario where only 10 - 20 \% of the gas is in a CNM.

A more plausible explanation for the conflicting results is that we cannot assume that a simple two-phase model consisting of a canonical CNM of T = 100 K and WNM of T = 8000 K is capable of reproducing the results for the large range of physical conditions applicable for all DLAs. Indeed considering the wide variety of ranges in metallicity, dust-to-gas ratios, and UV radiation fields, the results from both \citet{Wolfire1995} and \citet{Wolfe2003a} suggest that CNM temperatures can range from 10 K to 500 K, with higher temperatures more likely for lower metallicities, higher dust-to-gas ratios and higher UV radiation fields. From the {\siiistar} and {\ciistar} technique we can conclude that DLA Q0458$-$02 likely contains the bulk of the gas at a temperature of $\sim$ 300 K, still well within the range of a CNM phase as defined by \citet{Wolfe2003a}, and fully consistent with the result found in \citet{Kanekar2014}. DLA Q1157$+$014 is an exception as the results from this paper suggest it has the bulk of its gas at a temperature of $\sim$ 1000 K, which is inconsistent with a CNM or WNM, but again consistent with the temperature found in \citet{Kanekar2014}. It is likely the case that Q1157$+$014 is not representative of the DLA population as a whole as it has measurable levels of {\siiistar} (See Section \ref{disc:sum_os}). Furthermore, we would expect to find some DLAs with gas temperatures inconsistent with either the CNM or WNM as such gas is seen often in the local ISM \citep{Roy2013b}. 

\begin{figure}[b]
\epsscale{1.25}
\plotone{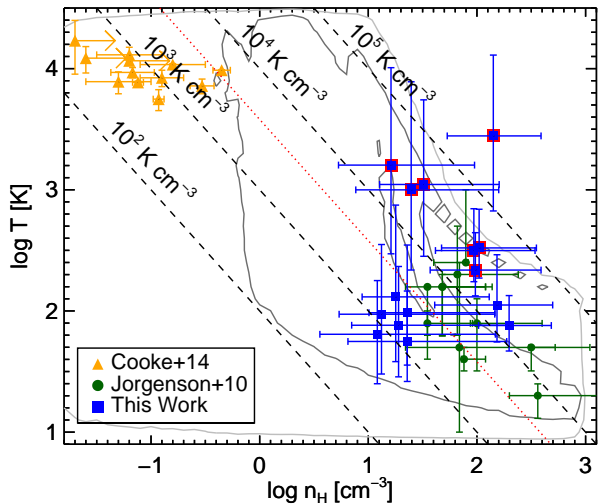}
\caption{Allowed temperature and neutral hydrogen density parameter space for DLAs. The larger (outlined in red) square data points are those with detectable levels of {\siiistar}. Overplotted on this figure are lines of constant pressure. The contours mark the 68 \% (dark gray) and 99 \% (light gray) confidence levels of the unbiased sample of \citet{Neeleman2013}. The dotted (red) line is the pressure of the local ISM as measured by \citet{Jenkins2011}.}
\label{fig:tvsnh2}
\end{figure} 

\subsubsection{\ion{C}{1} Fine-Structure Study}
A second method used to measure the physical parameters of DLAs is by considering the fine structure lines of \ion{C}{1}. This was done for several DLAs by \citet{Srianand2005} and \citet{Jorgenson2010}. \citet{Jorgenson2010} found that the densities and temperatures derived from this method could only result from very dense ({\nhi} $\ge$ 30 cm$^{-3}$) and cold gas (T $\le$ 150 K). They therefore surmise that \ion{C}{1} traces very dense pockets of very cold gas at slightly higher pressures. This is indeed seen in Figure \ref{fig:tvsnh2} where the results from \citet{Jorgenson2010} trace the coldest and densest measurements from our sample. Furthermore, the mean pressure from the \citet{Jorgenson2010} sample is higher than the median pressure for the complete sample in this paper ($\log(P/k_{\rm{B}})$ = 3.0 [K cm$^{-3}$]).

A comparison between the three DLAs that have been analyzed using both the \ion{C}{1} and the {\ciistar} and {\siiistar} analysis (Table \ref{tab:comp}) shows that both methods give remarkably similar temperatures and densities. This is somewhat at odds with the scenario put forth in \citet{Jorgenson2010}. They suggest that \ion{C}{1} traces small dense clumps of cold neutral gas in a larger less dense medium of cold gas. However, we find that for these three DLAs the \ion{C}{1} method gives values in agreement with the measurements of the bulk of the gas from the {\siiistar} and {\ciistar} technique, removing the need for this scenario in these DLAs.

One possible explanation for this result is that the \ion{C}{1} analysis can only be performed when multiple \ion{C}{1} fine-structure states can be measured. Such measurements are easiest for those DLAs with large column densities of the \ion{C}{1} fine-structure states, which results in preferentially selecting DLAs which contain the bulk of their gas in a cold and dense phase. This assessment is corroborated by the fact that 5 out of the 9 DLAs with 1-$\sigma$ temperature measurements below 500 K show \ion{C}{1} absorption. For the unbiased DLA sample of \citet{Neeleman2013}, the fraction of DLAs showing detectable levels of \ion{C}{1} is more than 10 times smaller; only 4 out of the 80 DLAs show \ion{C}{1} absorption. A positive detection of \ion{C}{1} is therefore a strong indicator that the DLA contains a significant fraction of cold, dense gas.

\subsubsection{Other Studies}
The third and final method discussed here for measuring the temperature of DLAs is the use of fitting routines to measure the Doppler parameter of individual components. By measuring a wide range of ion species, one is able to untangle the thermal broadening of the Doppler parameter from the turbulent or bulk motion of the gas. The thermal broadening gives an estimate of the temperature of the gas. This method has been used to find the temperature of individual components in DLAs, resulting in detection of both cold and warm gas \citep{Carswell2010, Carswell2012}. However, the multiple velocity components of a typical DLA, make this method daunting.

Recently this method has been used for a selection of very metal poor DLAs, which have simpler velocity structure \citep{Cooke2014, Dutta2014}. The result from these studies indicate that these DLAs have higher temperature and lower densities than the DLAs in our sample with conditions similar to those expected from a WNM. This suggests that the gas being traced by these very metal-poor DLAs is less likely to host star formation, which is corroborated by the lower metallicity of the gas.

Finally, we can compare our results to those found for the local ISM. Using \ion{C}{1}, \citet{Jenkins2011} find that the CNM in the local ISM has an average pressure of $\log(P/k_B)$ = 3.58$\pm$0.18 [K cm$^{-3}$]. The median pressure for our sample is $\log(P/k_B)$ = 3.0 [K cm$^{-3}$]. However, if we include only those DLAs with well-determined ranges on their pressure,  the median pressure becomes $\log(P/k_B)$ = 3.4 [K cm$^{-3}$]. We believe that this pressure is more representative of the complete DLA sample, as the large number of lower limits will artificially lower the median pressure. This pressure is very similar to the pressure found locally, although our sample has a larger range of allowed pressures. This extended range in pressures is easily explained by the fact that DLAs probe a variety of different galaxies, with a wider range of physical conditions compared to those seen in our own Galaxy.

\subsubsection{Summary}
\label{disc:sum_os}
Table \ref{tab:comp} lists the results for the 7 DLAs which were previously examined using either 21 cm absorption, \ion{C}{1} absorption or line profile fitting. These 7 DLAs show that for the limited sample of DLAs with temperature measurements from two different methods, the {\siiistar} and {\ciistar} method measures temperatures that are in general agreement with the results from other techniques. Two discrepancies exist. The temperature measurement for Q0458$-$02 from the 21 cm absorption study is likely high because the optical and radio line of sight encounter different column densities of gas \citep[see][]{Kanekar2014}. The only other measurement that is inconsistent within 1-$\sigma$ is that of DLA Q0336$-$01; this discrepancy is also discussed in \citet{Kanekar2014}. The remarkable agreement between the methods suggest that at least for the subset of DLAs with large fractions of cold gas, the {\siiistar} and {\ciistar} method is able to accurately determine the temperature and density of the gas.

The results from this section are summarized in Figure (Figure \ref{fig:tvsnh2}). The gray contours are the 68 \% and 99 \% confidence intervals of the unbiased sample of \citet{Neeleman2013}. The data points for our sample are those DLAs with well-defined ranges. Of these DLAs, the ones marked with larger squares (outlined in red) are those with measurable {\siiistar}. The DLAs with detectable levels of {\siiistar} fall outside the 68 \% contour intervals, indicating that the conditions conducive to {\siiistar} detections are not common in a random sample of DLAs. Indeed one DLA, J1135$-$0010, falls outside the 99 \% contour; this DLA, however, is unique in several other ways \citep[see][]{Kulkarni2012,Noterdaeme2012b} and therefore its physical conditions need not be similar to a typical DLA.

The measurements from the \ion{C}{1} method  by \citet{Jorgenson2010} are consistent with measurement for the coldest and densest DLAs in our sample. This is not unexpected as \ion{C}{1} likely traces the coldest gas in DLAs. On the other hand the metal poor sample of \citet{Cooke2014} have temperatures and densities consistent with a WNM. These DLAs fall outside the 68 \% contour of the unbiased sample, suggesting that the low densities for this sample are not common in a typical DLA and could be due to the very low metallicity of these DLAs. Finally, the dotted line in Figure \ref{fig:tvsnh2} is the average pressure of the local ISM \citep{Jenkins2011}, which is consistent with the pressures found in DLAs.

\subsection{Comparison with the Two-Phase Model}
\label{sec:disc_compmod}
As discussed in the introduction, \citet{Wolfe2003a} adopted the two-phase medium model from \citet{Wolfire1995} to describe the physical conditions of the gas around DLA galaxies. The results from this paper are able to test the validity of the two-phase model, since the {\siiistar} and {\ciistar} method provides independent measurements of the temperature and density of the DLA gas.

The first such test is to check that the two-phase model is able to reproduce the range of allowed pressures. We find allowable pressures ranges of $\log(P/k_{\rm{B}})$ between  1 [K cm$^{-3}$] and 6 [K cm$^{-3}$]. This large range of pressures is allowed within the two-phase model \citep[see Fig 5a and 5c of][]{Wolfe2003a}, since the lower metallicity of DLAs and varying star formation rate density can give rise to a large range of pressures that are able to maintain a stable two-phase structure.

A second test of the two-phase model is provided by comparing the star formation per unit area ($\Sigma_{\rm{SFR}}$) predicted from the two-phase model with that measured from emission lines of the DLA galaxy. Detecting DLA galaxies in emission is rare \citep[see e.g.][]{Krogager2012}; hence only 1 DLA (J1135$-$0010) in our sample has a published estimate for $\Sigma_{\rm{SFR}}$ from emission studies. We convert our pressure estimate and density measurement of this DLA into a star formation rate per unit area, in a similar way as was done in Figure 5 of \citet{Wolfe2003a}. Using this method we find a star formation rate of 0.3 M$_{\odot}$ yr$^{-1}$ kpc$^{-2}$. This compares well with the observed rate predicted from emission lines, which is $\sim$1 M$_{\odot}$ yr$^{-1}$ kpc$^{-2}$ \citep{Noterdaeme2012b}.

Finally, we can compare the temperatures from the {\siiistar} and {\ciistar} method, to see if we find any evidence for two distinct phases, which is a prediction of the two-phase model. As discussed in Section \ref{sec:disc_compdat}, we find that at least 5 \% of DLAs have a significant fraction of cold gas, consistent with a canonical CNM (Figure \ref{fig:tvsnh}). Unfortunately, we are not able to confirm the existence of gas in a WNM as our method provides weak constraints at high temperatures. However, from other studies such as \citet{Carswell2012} and \citet{Cooke2014}, we know that such gas exists. 

In conclusion, the results in this paper are in general agreement with the two-phase model of \citet{Wolfe2003a, Wolfe2003b}. There are several DLAs, however, which have higher than predicted temperatures and densities; these DLAs are discussed further in Section \ref{sec:disc_impl}. 

\subsection{Implications for High-$z$ Galaxies}
\label{sec:disc_impl}
In this section we will speculate about the implications these results have on the physical conditions of DLA gas and the implication on the formation of high-$z$ galaxies. 

\subsubsection{Implications for DLA Gas}
In this paper we have focussed on the physical conditions of the bulk of the neutral gas for a large sample of DLAs. This is unlike previous absorption studies using \ion{C}{1}, which focus solely on the coldest and densest gas of DLAs as was noted by \citet{Jorgenson2010}. The results from this paper corroborates this assessment as the results from the \ion{C}{1} analysis are consistent with the coldest and densest gas measurements from the {\siiistar} and {\ciistar} method.

We find that the fraction of DLAs which have their bulk of gas in such a cold and dense phase must be at least 5 \%. For the remaining DLAs the amount of gas in such a phase is unknown as the Si$^+$ and C$^+$ ratios do not provide stringent constraints on the temperature and density of the gas. Interestingly, we find that the upper levels of both the ground state of Si$^+$ and C$^+$ have the same velocity structure as the lower levels for the majority of DLAs, suggesting similar conditions for the majority of the velocity components in these DLAs.

\begin{figure}[t]
\epsscale{1.2}
\plotone{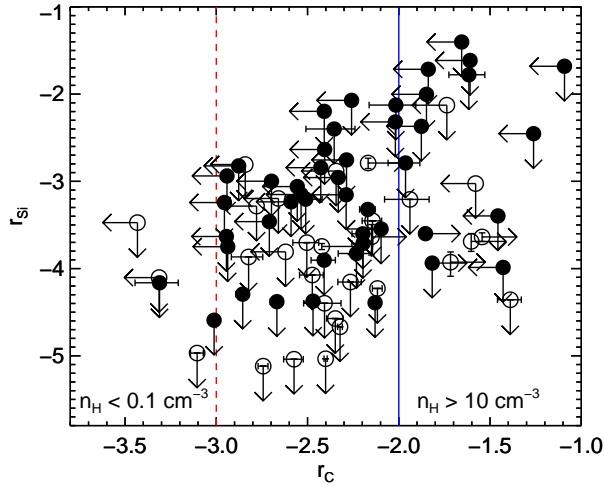}
\caption{{\rc} and {\rsi} ratio for the complete sample. The filled circle are the 46 DLAs that are part of the unbiased sample of \citet{Neeleman2013} and have a well-defined {\rc} and {\rsi}. The region to the left of the dashed (red) line can only be occupied by DLAs with neutral hydrogen density less than 0.1 cm$^{-3}$, whereas the region to the left of the solid (blue) line can be occupied solely by DLAs with neutral hydrogen densities greater than 10 cm$^{-3}$.}
\label{fig:rcrsi}
\end{figure} 

The range of allowed temperatures and densities is summarized in Figure \ref{fig:tvsnh2}. The unbiased sample covers the parameter space between the metal-poor sample and the sample of \ion{C}{1} detections. Several DLAs have temperature and density measurements which are inconsistent with the two-phase model of \citet{Wolfe2003a}. The enhanced densities and temperatures in these DLAs increases {\rc} (see Figure \ref{fig:sicmod}), and therefore the observed cooling rate, {\elc}. As a result, setting the observed cooling rate equal to the calculated cooling rate from a two-phase model will over predict the star formation rate for these DLAs.

This could partially provide the answer to the unsuccessful attempts of observing the DLAs in emission \citep{Fumagalli2014}. If a large fraction of DLAs have enhanced densities and temperatures compared to the two-phase model, the average star formation rate for DLAs will be systematically overestimated. One possible explanation for the enhanced temperatures and pressures could be turbulence. At least in numerical simulations turbulence is able to drive some of the gas into the classically forbidden region \citep{Gazol2005}. Such gas will have enhanced {\rc} ratios compared to those predicted from the two-phase model and therefore the two phase model will overestimate the star formation rate.

In conclusion, we suggest that the gas in DLAs follows in general the two-phase model of \citet{Wolfe2003a}, as several studies have measured gas with properties very similar to both the WNM \citep{Lehner2008,Carswell2012,Kanekar2014,Cooke2014} and CNM \citep{Howk2005,Srianand2005,Carswell2010, Jorgenson2010}. However, the detection of {\siiistar} in DLAs suggest that a fraction of DLAs have significantly higher densities and temperatures than expected from the two-phase model. For these systems the star formation rate obtained from the two-phase model is overestimated. Unfortunately, the exact fraction of these DLAs cannot be estimated from this study. However, in the local ISM this fraction is quite significant ($\sim$ 30 \%) \citep{Roy2013b}. 

We would like to note that these higher temperatures are in agreement with the results from 21 cm absorption, thereby resolving two problems plaguing DLA studies at once; the lack of detections of DLAs in emission and the discrepancy between temperatures expected from the two-phase model and measurements from 21 cm studies. One possible adaptation to the two-phase model which could provide such a solution is to include turbulence, as numerical simulations show that turbulence could drive gas into the classically forbidden region.

\subsubsection{Implications for High-$z$ Galaxy Formation}
\label{sec:disc_implgal}
By the nature of their selection, DLA sightlines represent a cross-section weighted sampling of high-surface density, neutral
hydrogen gas at high-$z$. In aggregate, these systems also represent a major reservoir to fuel galaxy formation during the first few Gyr of the universe \citep{Wolfe1995, Prochaska2005}.   Therefore, one generally associates this gas with the ISM of young galaxies.  As such, the results presented here offer new insight into the nature of this ISM gas and its relationship to ongoing or future star-formation.

Restricting the discussion first to our random sample, we find that the incidence of very strong fine-structure absorption is rare:  {\rc} exceeds $10^{-2}$ in only 4 out of the 46 DLAs from the random sample with measured {\rc} (see Figure \ref{fig:rcrsi}). Such high {\rc} values require gas densities {\nhi} $\gtrsim$ 10 cm$^{-3}$ for the majority of neutral, atomic gas at high-$z$. In conjunction with the paucity of systems showing molecular gas and/or \ion{C}{1} detection, which are both indicators of dense gas, these results suggest that a large fraction of DLA gas is unconducive to star-formation. A result corroborated by the difficulty to directly measure the in-situ star-formation of a typical DLA \citep[see e.g.][]{Fumagalli2014}. Indeed, this material may even form so-called `dark galaxies' \citep{Cantalupo2012}. 

A significant fraction of DLAs (9 out 46) have measured {\rc} value in the range $10^{-2.5}$ to $10^{-2}$. For the single-phase analysis performed in this manuscript, we derive {\nhi} $\gtrsim 3 $ cm $^{-3}$ which exceeds the canonical value for the WNM. Together with the 18 DLAs which have upper limits to {\rc} that exceed $10^{-2.5}$, about half of all systems are consistent with moderate densities which exceed those expected in a canonical WNM. One might argue, however, that the systems with intermediate {\rc} values represent a mixture of dense and more diffuse gas with the dense gas contributing nearly all of the observed {\ciistar} absorption. Then, the bulk of the gas could (in principle)  be very diffuse. However, as mentioned in Section \ref{sec:tech}, we disfavor extreme scenarios of this kind because the absorption profiles of {\lambdaciistar} so closely track the resonance lines, in velocity and optical depth.  This shows that there are no substantial regions along the sightline of highly diffuse gas without corresponding dense gas.  Furthermore, even if one adopts a two-phase medium in pressure equilibrium with the WNM  dominating the column density, then the {\nhi} value derived from a single-phase analysis only overestimates the mass-weighted value by a factor of a few.

Despite a large fraction of DLAs favoring modest densities, the majority of the DLAs have significant gas pressures ($P > 10^3$ K cm$^{-3}$), which is a characteristic of an active ISM.  Recent models of galaxy formation within hierarchical cosmology predict highly turbulent conditions driven by the accretion of cool gas and violent disk instabilities within the galaxies 
\citep[e.g.][]{Keres2005,Dekel2006,Burkert2010,Fumagalli2011}. Perhaps such processes explain the small, but non-negligible subset of DLA sightlines with $P> 10^4$ K cm$^{-3}$. As noted in Section \ref{sec:disc_data}, these pressures are predominantly recorded in gas with high metallicity and large velocity widths.

\section{Summary}
\label{sec:summ}
In this paper we have presented a new method of determining the physical conditions of gas in high redshift galaxies. Using the fine-structure lines of Si$^+$ and C$^+$, we are able to provide constraints on the temperature and neutral hydrogen density of DLAs. We have applied this method to a sample of 80 DLAs, for which we are able to provide limits or detections of these fine-structure line transitions. This sample contains 5 new detections of the excited fine-structure line of Si$^+$, which more than doubles the previously know detections. The results of this analysis are:

\begin{enumerate}
\item
We find that 9 DLAs have temperatures consistent with gas in a cold neutral medium. The remaining DLAs provide less stringent constraints on their temperature for two reasons. Firstly, the ratios of fine-structure lines become insensitive to temperature changes above 500 K. Secondly, the low density of {\siiistar} and {\ciistar} in these systems makes detection difficult; resulting in weak upper limits to the column density measurements of both fine-structure lines in these systems.

\item
From the `unbiased' subsample of DLAs part of the sample described in \citet{Neeleman2013}, we find that at least 5 \% of all  DLAs have significant fractions of gas with properties similar to a canonical CNM along their line-of-sight. This result is consistent with the locally measured volume filling fraction of 0.01 for the CNM. 

\item
The results of the method show that the neutral hydrogen density of DLAs vary significantly from DLA to DLA. On the other hand the electron density varies little between DLAs with a median electron density of $0.0044 \pm 0.0028~\rm{cm}^{-3}$. Furthermore, we can rule out the parameter space of high temperature and high neutral hydrogen density (see Figure \ref{fig:tvsnh}) as such gas would produce upper to lower level fine-structure state ratios in excess of what we observe.

\item
We find that there exist a correlation between the neutral hydrogen density and the cooling rate of the DLA. This is consistent with the predictions from the two phase model, where stronger star formation rates and therefore larger cooling rates result in higher stable neutral hydrogen equilibrium densities. Furthermore, the comparison between the neutral hydrogen density and the total {\hi} column density gives a rough estimate of the total absorption length along the line-of-sight. These values range from about 1 kpc to only a few pc, suggesting that the bulk of the neutral gas at high redshift can be located in reasonably small dense components. 

\item
Finally, we find that the typical pressure of the DLAs in the sample is $\log(P/k_B)$ = 3.4 [K cm$^{-3}$], which is comparable to the pressure of the local ISM. However, the DLAs show a larger range in pressures, which can be easily explained by the fact that DLAs measure a range of different galaxies, with a wide range of different physical conditions.

\end{enumerate}

We speculate that these results indicate that DLAs generally follow the two-phase model of \citet{Wolfe2003b}. However, a fraction of DLAs have temperatures and densities inconsistent with this model. As a result, the two-phase model will over predict the star formation rate of these systems. By including a mechanism in the two-phase model which will increase the temperature and density of the gas for these DLAs, we can account for both the higher spin temperatures seen in 21 cm absorption \citep{Kanekar2014}, and lower the star formation rates of DLAs as is suggested by recent observations \citep{Fumagalli2014}. One such mechanism is turbulence, which is able to drive gas into the unstable temperature regime \citep{Gazol2005,Walch2011}.

\section{Acknowledgements}
\label{sec:ack}
We wish to thank the referee for the helpful comments and N. Kanekar for reading a previous version of the manuscript. Support for this work was provided by NSF award AST-1109452. The data presented herein were obtained at the W.M. Keck Observatory, which is operated as a scientific partnership among the California Institute of Technology, the University of California and the National Aeronautics and Space Administration. The Observatory was made possible by the generous financial support of the W.M. Keck Foundation. The authors wish to recognize and acknowledge the very significant cultural role and reverence that the summit of Mauna Kea has always had within the indigenous Hawaiian community. We are most fortunate to have the opportunity to conduct observations from this mountain.

\bibliography{Bib}

\newpage
\LongTables
\begin{deluxetable*}{llccccccccl}
\tabletypesize{\scriptsize}
\tablenum{1}
%\rotate
\tablecaption{{Fine Structure DLA sample}
\label{tab:meas}}
\tablewidth{0pt}
\tablehead{
\colhead{Index} &
\colhead{QSO} &
\colhead{$z_{\rm{abs}}$} &
\colhead{$\log N_{\rm HI}$} &
\colhead{Metallicity} &
\colhead{M\,\tablenotemark{a}} &
\colhead{${\log} N_{\rm CII*}$} &
\colhead{${\log} N_{\rm SiII*}$} &
\colhead{${\log} N_{\rm SiII}$} &
\colhead{N13\,\tablenotemark{b}} &
\colhead{References} \\
\colhead{Number} &
\colhead{} &
\colhead{} &
\colhead{(cm$^{-2}$)} &
\colhead{[M/H]} &
\colhead{} &
\colhead{(cm$^{-2}$)} &
\colhead{(cm$^{-2}$)} &
\colhead{(cm$^{-2}$)} &
\colhead{} &
\colhead{} \\
}
\startdata
1 & Q1157$+$014 & 1.9437 & 21.70 $\pm$ 0.10 & $-1.23$ $\pm$ 0.10 & Si & $>$ $14.75$\,\tablenotemark{c} & $12.37$$\pm$0.03 & $15.97$$\pm$0.01 & Y & 5, 18\\
2 & Q1215$+$33 & 1.9991 & 20.95 $\pm$ 0.06 & $-1.43$ $\pm$ 0.07 & Si & $<$ $13.18$ & $<$ $12.56$ & $15.02$$\pm$0.02 & Y & 4, 7\\
3 & Q0458$-$02 & 2.0395 & 21.65 $\pm$ 0.09 & $-1.11$ $\pm$ 0.09 & Si & $>$ $14.88$\,\tablenotemark{c} & $<$ $12.58$ & $16.04$$\pm$0.02\,\tablenotemark{c} & Y & 4, 7\\
4 & J2340$-$0053 & 2.0545 & 20.35 $\pm$ 0.15 & $-0.55$ $\pm$ 0.15 & S & $13.72$$\pm$0.01 & $<$ $11.31$ & $15.23$$\pm$0.01 & Y & 12, 17, 19\\
5 & Q2206$-$19 & 2.0762 & 20.43 $\pm$ 0.06 & $-2.25$ $\pm$ 0.07 & Si & $<$ $13.20$ & $<$ $11.83$ & $13.68$$\pm$0.03 & Y & 2, 4, 7\\
6 & Q2359$-$02 & 2.0950 & 20.70 $\pm$ 0.10 & $-0.72$ $\pm$ 0.10 & Si & $13.69$$\pm$0.06 & $<$ $12.05$ & $15.48$$\pm$0.02 & Y & 4, 7\\
7 & Q0149$+$33 & 2.1407 & 20.50 $\pm$ 0.10 & $-1.43$ $\pm$ 0.11 & Si & $<$ $12.79$ & $<$ $12.22$ & $14.57$$\pm$0.04 & Y & 4, 7\\
8 & Q2348$-$14 & 2.2794 & 20.56 $\pm$ 0.07 & $-2.01$ $\pm$ 0.15 & S & $<$ $13.19$ & $<$ $12.04$ & $14.20$$\pm$0.02 & Y & 4, 7\\
9 & J2036$-$0553 & 2.2805 & 21.20 $\pm$ 0.15 & $-1.81$ $\pm$ 0.15 & S & $13.41$$\pm$0.03 & $<$ $11.86$ & $14.99$$\pm$0.04 & Y & 18\\
10 & J1435$+$5359 & 2.3427 & 21.05 $\pm$ 0.10 & $-1.42$ $\pm$ 0.10 & S & $12.88$$\pm$0.03 & $<$ $11.30$ & $15.12$$\pm$0.01 & Y & 16\\
11 & Q2343$+$125 & 2.4313 & 20.34 $\pm$ 0.10 & $-0.47$ $\pm$ 0.10 & Si & $12.68$$\pm$0.11 & $<$ $11.69$ & $15.37$$\pm$0.01 & Y & 3, 11\\
12 & J1541$+$3153 & 2.4435 & 20.95 $\pm$ 0.10 & $-1.49$ $\pm$ 0.10 & Si & $13.07$$\pm$0.09 & $<$ $12.23$ & $14.96$$\pm$0.03 & Y & 21\\
13 & Q0836$+$11 & 2.4652 & 20.60 $\pm$ 0.10 & $-1.11$ $\pm$ 0.11 & Si & $<$ $13.13$ & $<$ $12.23$ & $14.99$$\pm$0.04 & Y & 7\\
14 & Q2344$+$12 & 2.5379 & 20.36 $\pm$ 0.10 & $-1.69$ $\pm$ 0.10 & Si & $<$ $12.84$ & $<$ $11.96$ & $14.18$$\pm$0.01 & Y & 1, 7\\
15 & Q0913$+$072 & 2.6184 & 20.35 $\pm$ 0.10 & $-2.52$ $\pm$ 0.10 & Si & $<$ $12.77$ & $<$ $12.41$ & $13.33$$\pm$0.02 & Y & 13\\
16 & Q1759$+$75 & 2.6253 & 20.80 $\pm$ 0.10 & $-0.70$ $\pm$ 0.10 & S & $13.14$$\pm$0.03 & $<$ $11.60$ & $15.53$$\pm$0.01 & Y & 4, 7\\
17 & J1035$+$5440 & 2.6840 & 20.50 $\pm$ 0.20 & $-0.94$ $\pm$ 0.20 & Zn & $>$ $13.21$ & $<$ $11.69$ & $>$ $15.10$ & Y & 18\\
18 & PKS1354$-$17 & 2.7800 & 20.30 $\pm$ 0.15 & $-1.31$ $\pm$ 0.18 & Si & $12.75$$\pm$0.06 & $<$ $12.56$ & $14.49$$\pm$0.11 & Y & 10, 18\\
19 & HS1132$+$2243 & 2.7834 & 21.00 $\pm$ 0.07 & $-1.90$ $\pm$ 0.07 & S & $<$ $12.69$ & $<$ $11.42$ & $>$ $14.48$ & Y & 10\\
20 & Q1337$+$11 & 2.7958 & 20.95 $\pm$ 0.10 & $-1.75$ $\pm$ 0.10 & S & $<$ $13.11$ & $<$ $11.63$ & $14.78$$\pm$0.06 & Y & 10, 17\\
21 & J1353$+$5328 & 2.8349 & 20.80 $\pm$ 0.10 & $-1.37$ $\pm$ 0.10 & S & $13.20$$\pm$0.05 & $<$ $11.46$ & $14.81$$\pm$0.06 & Y & 21\\
22 & J1131$+$6044 & 2.8757 & 20.50 $\pm$ 0.15 & $-1.80$ $\pm$ 0.15 & O & $<$ $12.73$ & $<$ $11.64$ & $13.92$$\pm$0.01 & Y & 18\\
23 & J1304$+$1202 & 2.9131 & 20.55 $\pm$ 0.15 & $-1.65$ $\pm$ 0.15 & S & $<$ $12.81$ & $<$ $11.58$ & $14.25$$\pm$0.04 & Y & 21\\
24 & J1304$+$1202 & 2.9289 & 20.30 $\pm$ 0.15 & $-1.54$ $\pm$ 0.15 & S & $<$ $12.77$ & $<$ $11.66$ & $>$ $13.85$ & Y & 21\\
25 & Q1021$+$30 & 2.9489 & 20.70 $\pm$ 0.10 & $-1.97$ $\pm$ 0.12 & S & $<$ $12.76$ & $<$ $11.60$ & $14.37$$\pm$0.08 & Y & 7, 10\\
26 & J1014$+$4300 & 2.9588 & 20.50 $\pm$ 0.10 & $-0.83$ $\pm$ 0.10 & Si & $12.63$$\pm$0.06 & $<$ $11.82$ & $15.17$$\pm$0.01 & Y & 18\\
27 & J1410$+$5111 & 2.9642 & 20.85 $\pm$ 0.20 & $-1.94$ $\pm$ 0.15 & Si & $<$ $13.01$ & $<$ $11.78$ & $14.34$$\pm$0.02 & Y & 18\\
28 & HS0741$+$4741 & 3.0174 & 20.40 $\pm$ 0.10 & $-1.54$ $\pm$ 0.10 & S & $<$ $12.51$ & $<$ $11.08$ & $14.35$$\pm$0.01 & Y & 7\\
29 & J1240$+$1455 & 3.0241 & 20.45 $\pm$ 0.10 & $-1.39$ $\pm$ 0.13 & S & $<$ $13.31$ & $<$ $12.54$ & $>$ $14.37$ & Y & 18\\
30 & Q0347$-$38 & 3.0247 & 20.60 $\pm$ 0.10 & $-1.08$ $\pm$ 0.10 & Si & $13.47$$\pm$0.03 & $<$ $12.17$ & $15.02$$\pm$0.02 & Y & 4, 7, 8\\
31 & Q0336$-$01 & 3.0620 & 21.20 $\pm$ 0.10 & $-1.49$ $\pm$ 0.10 & S & $14.00$$\pm$0.01 & $<$ $11.65$ & $>$ $14.87$ & Y & 7\\
32 & J1200$+$4015 & 3.2200 & 20.85 $\pm$ 0.10 & $-0.64$ $\pm$ 0.10 & S & $13.68$$\pm$0.02 & $<$ $11.84$ & $>$ $15.21$ & Y & 21\\
33 & Q0930$+$28 & 3.2352 & 20.30 $\pm$ 0.10 & $-1.69$ $\pm$ 0.11 & O & $<$ $12.57$ & $<$ $12.16$ & $13.88$$\pm$0.02 & Y & 9\\
34 & J0900$+$4215 & 3.2458 & 20.30 $\pm$ 0.10 & $-0.82$ $\pm$ 0.10 & S & $13.07$$\pm$0.03 & $<$ $11.23$ & $15.11$$\pm$0.01 & Y & 17, 16\\
35 & J0929$+$2825 & 3.2627 & 21.10 $\pm$ 0.10 & $-1.56$ $\pm$ 0.10 & S & $13.19$$\pm$0.02 & $<$ $11.26$ & $15.04$$\pm$0.01 & Y & 18\\
36 & J2315$+$1456 & 3.2732 & 20.30 $\pm$ 0.15 & $-1.68$ $\pm$ 0.16 & S & $<$ $12.88$ & $<$ $11.64$ & $14.12$$\pm$0.01 & Y & 18\\
37 & BR0019$-$15 & 3.4388 & 20.92 $\pm$ 0.10 & $-1.01$ $\pm$ 0.11 & Si & $13.84$$\pm$0.01 & $<$ $12.41$ & $15.41$$\pm$0.05 & Y & 4, 7\\
38 & J0814$+$5029 & 3.7075 & 21.35 $\pm$ 0.15 & $-2.07$ $\pm$ 0.15 & S & $13.09$$\pm$0.04 & $<$ $12.12$ & $14.43$$\pm$0.07 & Y & 18\\
39 & BRI1346$-$03 & 3.7358 & 20.72 $\pm$ 0.10 & $-2.27$ $\pm$ 0.10 & Si & $12.55$$\pm$0.12 & $<$ $12.30$ & $13.95$$\pm$0.01 & Y & 4, 7\\
40 & PSS0209$+$05 & 3.8635 & 20.55 $\pm$ 0.10 & $-2.54$ $\pm$ 0.10 & Si & $<$ $12.51$ & $<$ $11.91$ & $13.51$$\pm$0.01 & Y & 10\\
41 & J1051$+$3107 & 4.1392 & 20.70 $\pm$ 0.20 & $-1.99$ $\pm$ 0.21 & S & $<$ $13.01$ & $<$ $11.52$ & $14.20$$\pm$0.01 & Y & 21\\
42 & PSS1443$+$27 & 4.2241 & 21.00 $\pm$ 0.10 & $-0.94$ $\pm$ 0.17 & Fe & $<$ $14.68$ & $<$ $11.51$ & $>$ $15.32$ & Y & 6, 7\\
43 & J0817$+$1351 & 4.2584 & 21.30 $\pm$ 0.15 & $-1.15$ $\pm$ 0.15 & S & $>$ $14.40$ & $<$ $11.83$ & $>$ $14.93$ & Y & 21\\
44 & J1100$+$1122 & 4.3947 & 21.74 $\pm$ 0.10 & $-1.68$ $\pm$ 0.21 & Fe & $14.15$$\pm$0.02 & $<$ $12.57$ & $>$ $14.85$ & Y & 21\\
45 & J1607$+$1604 & 4.4741 & 20.30 $\pm$ 0.15 & $-1.70$ $\pm$ 0.15 & Si & $13.10$$\pm$0.09 & $<$ $12.80$ & $14.10$$\pm$0.01 & Y & 21\\
46 & J1200$+$4618 & 4.4765 & 20.50 $\pm$ 0.15 & $-1.38$ $\pm$ 0.22 & Fe & $<$ $13.73$ & $<$ $13.11$ & $>$ $14.55$ & Y & 21\\
47 & J1202$+$3235 & 4.7955 & 21.10 $\pm$ 0.15 & $-2.35$ $\pm$ 0.22 & Fe & $13.29$$\pm$0.02 & $<$ $12.30$ & $>$ $13.80$ & Y & 21\\
48 & J1051$+$3545 & 4.8206 & 20.35 $\pm$ 0.10 & $-2.27$ $\pm$ 0.10 & Si & $<$ $12.83$ & $<$ $12.34$ & $13.58$$\pm$0.02 & Y & 21\\
49 & J1056$+$1208 & 1.6093 & 21.45 $\pm$ 0.15 & $-0.47$ $\pm$ 0.17 & Si & $>$ $15.65$\,\tablenotemark{c} & $12.79$$\pm$0.04 & $16.48$$\pm$0.08\,\tablenotemark{c} & N & 20, 24, 25\\
50 & J0044$+$0018 & 1.7250 & 20.35 $\pm$ 0.10 & $-0.23$ $\pm$ 0.10 & S & $13.81$$\pm$0.03 & $<$ $12.36$ & $15.34$$\pm$0.04 & N & 20, 25\\
51 & J0927$+$1543 & 1.7311 & 21.35 $\pm$ 0.15 & $-0.86$ $\pm$ 0.15 & Si & $>$ $15.22$\,\tablenotemark{c} & $<$ $12.11$ & $15.99$$\pm$0.01 & N & 25\\
52 & J0008$-$0958 & 1.7675 & 20.85 $\pm$ 0.15 & $-0.16$ $\pm$ 0.15 & S & $14.34$$\pm$0.01 & $<$ $11.88$ & $16.04$$\pm$0.01 & N & 15, 25\\
53 & J1249$-$0233 & 1.7808 & 21.45 $\pm$ 0.15 & $-1.06$ $\pm$ 0.15 & S & $14.12$$\pm$0.01 & $<$ $12.26$ & $>$ $15.11$ & N & 15, 25\\
54 & J0233$+$0103 & 1.7850 & 20.60 $\pm$ 0.15 & $-1.34$ $\pm$ 0.15 & Si & $<$ $13.03$ & $<$ $12.44$ & $14.77$$\pm$0.09 & N & 25\\
55 & J1454$+$0941 & 1.7884 & 20.50 $\pm$ 0.15 & $-0.39$ $\pm$ 0.16 & S & $13.59$$\pm$0.06 & $<$ $12.24$ & $15.47$$\pm$0.03 & N & 25\\
56 & J1313$+$1441 & 1.7947 & 21.20 $\pm$ 0.15 & $-0.59$ $\pm$ 0.15 & Si & $>$ $15.26$\,\tablenotemark{c} & $12.47$$\pm$0.02 & $16.11$$\pm$0.04\,\tablenotemark{c} & N & 24, 25\\
57 & J1310$+$5424 & 1.8005 & 21.45 $\pm$ 0.15 & $-0.51$ $\pm$ 0.15 & Si & $>$ $15.43$\,\tablenotemark{c} & $12.51$$\pm$0.12 & $16.44$$\pm$0.04\,\tablenotemark{c} & N & 20, 24, 25\\
58 & J1106$+$1044 & 1.8185 & 20.50 $\pm$ 0.15 & $-0.32$ $\pm$ 0.15 & S & $<$ $13.36$ & $<$ $12.19$ & $>$ $15.21$ & N & 25\\
59 & J1142$+$0701 & 1.8407 & 21.50 $\pm$ 0.15 & $-0.85$ $\pm$ 0.20 & Si & $>$ $14.83$\,\tablenotemark{c} & $12.51$$\pm$0.09 & $16.15$$\pm$0.13\,\tablenotemark{c} & N & 24, 25\\
60 & J0815$+$1037 & 1.8462 & 20.30 $\pm$ 0.15 & $-0.43$ $\pm$ 0.47 & Si & $<$ $13.69$ & $<$ $12.57$ & $15.37$$\pm$0.44 & N & 25\\
61 & J1335$+$0824 & 1.8560 & 20.65 $\pm$ 0.15 & $-0.51$ $\pm$ 0.15 & S & $13.87$$\pm$0.05 & $<$ $12.12$ & $15.72$$\pm$0.03 & N & 25\\
62 & J1024$+$0600 & 1.8950 & 20.60 $\pm$ 0.15 & $-0.30$ $\pm$ 0.15 & S & $14.31$$\pm$0.01 & $<$ $12.05$ & $15.80$$\pm$0.02 & N & 25\\
63 & J1524$+$1030 & 1.9409 & 21.65 $\pm$ 0.15 & $-0.75$ $\pm$ 0.15 & Zn & $>$ $14.76$\,\tablenotemark{c} & $<$ $11.55$ & $>$ $16.24$\,\tablenotemark{c} & N & 24, 25\\
64 & J1042$+$0628 & 1.9429 & 20.70 $\pm$ 0.15 & $-0.77$ $\pm$ 0.15 & S & $<$ $14.15$ & $<$ $12.06$ & $15.40$$\pm$0.07 & N & 25\\
65 & J1417$+$4132 & 1.9508 & 21.85 $\pm$ 0.15 & $-0.93$ $\pm$ 0.15 & Zn & $>$ $15.11$\,\tablenotemark{c} & $12.43$$\pm$0.03 & $>$ $16.41$\,\tablenotemark{c} & N & 23, 24, 25\\
66 & J1552$+$4910 & 1.9599 & 21.15 $\pm$ 0.15 & $-0.96$ $\pm$ 0.15 & S & $13.49$$\pm$0.03 & $<$ $11.49$ & $15.98$$\pm$0.01 & N & 25\\
67 & Q1755$+$578 & 1.9692 & 21.40 $\pm$ 0.15 & $-0.18$ $\pm$ 0.15 & Zn & $>$ $14.77$\,\tablenotemark{c} & $12.83$$\pm$0.03 & $16.57$$\pm$0.01\,\tablenotemark{c} & N & 24, 25\\
68 & J1305$+$0924 & 2.0184 & 20.40 $\pm$ 0.15 & $-0.15$ $\pm$ 0.15 & S & $14.10$$\pm$0.04 & $<$ $12.07$ & $15.75$$\pm$0.04 & N & 25\\
69 & J1509$+$1113 & 2.0283 & 21.30 $\pm$ 0.15 & $-0.76$ $\pm$ 0.15 & S & $14.31$$\pm$0.03 & $<$ $11.94$ & $16.04$$\pm$0.02 & N & 25\\
70 & J1135$-$0010 & 2.2068 & 22.05 $\pm$ 0.10 & $-1.07$ $\pm$ 0.10 & Si & $>$ $15.00$\,\tablenotemark{c} & $13.70$$\pm$0.05 & $16.49$$\pm$0.03\,\tablenotemark{c} & N & 22, 24\\
71 & J1211$+$0422 & 2.3765 & 20.70 $\pm$ 0.10 & $-1.22$ $\pm$ 0.10 & S & $<$ $12.73$ & $<$ $11.90$ & $14.94$$\pm$0.04 & N & 7, 16\\
72 & J2241$+$1225 & 2.4179 & 21.15 $\pm$ 0.10 & $-1.28$ $\pm$ 0.20 & Fe & $13.39$$\pm$0.09 & $<$ $12.72$ & $>$ $14.67$ & N & 25\\
73 & J0211$+$1241 & 2.5951 & 20.60 $\pm$ 0.15 & $-0.57$ $\pm$ 0.16 & Si & $13.32$$\pm$0.03 & $<$ $12.14$ & $15.53$$\pm$0.07 & N & 25\\
74 & J0812$+$3208 & 2.6263 & 21.35 $\pm$ 0.10 & $-0.55$ $\pm$ 0.13 & O & $14.02$$\pm$0.01 & $<$ $11.78$ & $15.98$$\pm$0.05 & N & 10, 17, 16, 19\\
75 & J1558$-$0031 & 2.7026 & 20.67 $\pm$ 0.05 & $-1.74$ $\pm$ 0.05 & S & $<$ $12.60$ & $<$ $11.54$ & $>$ $14.15$ & N & 14\\
76 & FJ2334$-$09 & 3.0569 & 20.45 $\pm$ 0.10 & $-0.99$ $\pm$ 0.10 & Si & $<$ $12.81$ & $<$ $11.97$ & $14.96$$\pm$0.03 & N & 10\\
77 & J2100$-$0641 & 3.0924 & 21.05 $\pm$ 0.15 & $-0.70$ $\pm$ 0.15 & S & $14.09$$\pm$0.01 & $<$ $11.31$ & $15.87$$\pm$0.01 & N & 15, 19, 20\\
78 & J1155$+$0530 & 3.3260 & 21.05 $\pm$ 0.10 & $-0.79$ $\pm$ 0.10 & S & $13.81$$\pm$0.02 & $<$ $11.29$ & $15.94$$\pm$0.01 & N & 18\\
79 & J0825$+$3544 & 3.6567 & 21.25 $\pm$ 0.10 & $-1.83$ $\pm$ 0.13 & Si & $13.14$$\pm$0.05 & $<$ $11.00$ & $14.92$$\pm$0.08 & N & 21\\
80 & J0909$+$3303 & 3.6581 & 20.55 $\pm$ 0.10 & $-0.89$ $\pm$ 0.10 & S & $13.53$$\pm$0.08 & $<$ $12.50$ & $14.85$$\pm$0.08 & N & 21
\enddata
\tablenotetext{a}{Ion used for metallicity determination}
\tablenotetext{b}{Part of the \citet{Neeleman2013} sample}
\tablenotetext{c}{VPFIT used to determine the column density}
\tablerefs{(1) \citet{Lu1996};(2) \citet{Prochaska1997};(3) \citet{Lu1998};(4) \citet{Prochaska1999};(5) \citet{Petitjean2000};(6) \citet{Prochaska2000};(7) \citet{Prochaska2001};(8) \citet{Levshakov2002};(9) \citet{Prochaska2002a};(10) \citet{Prochaska2003a};(11) \citet{Dessauges2004};(12) \citet{Khare2004};(13) \citet{Ledoux2006};(14) \citet{OMeara2006};(15) \citet{Herbert-Fort2006};(16) \citet{Dessauges2007};(17) \citet{Prochaska2007};(18) \citet{Wolfe2008};(19) \citet{Jorgenson2010};(20) \citet{Kaplan2010};(21) \citet{Rafelski2012};(22) \citet{Kulkarni2012};(23) \citet{Berg2013};(24) This Work;(25) \citet{Berg2014}}
\end{deluxetable*}

\newpage
\LongTables
\begin{deluxetable*}{llcccccccl}
\tabletypesize{\scriptsize}
\tablenum{2}
%\rotate
\tablecaption{{Results of {\siiistar} and {\ciistar} Technique}
\label{tab:res}}
\tablewidth{0pt}
\tablehead{
\colhead{Index} &
\colhead{QSO} &
\colhead{$z_{\rm{abs}}$} &
\colhead{$\log$ {\nhi}(2-$\sigma$)} &
\colhead{$\log$ {\nel}(2-$\sigma$)} &
\colhead{$\log$ T(2-$\sigma$)} &
\colhead{$\log$ P/k$_{\rm{B}}$(2-$\sigma$)} \\
\colhead{Number} &
\colhead{} &
\colhead{} &
\colhead{[cm$^{-3}$]} &
\colhead{[cm$^{-3}$]} &
\colhead{[K]} &
\colhead{[K cm$^{-3}$]} \\
}
\startdata
1 & Q1157$+$014    & 1.9437 & $1.1$-$2.2(0.9$-$2.8)$ & $-2.6$-$-1.2(-2.9$-$-1.0)$ & $2.5$-$3.7(2.2$-$4.1)$ & $ 3.8$-$5.5( 3.5$-$6.3)$ & \\
2 & Q1215$+$33     & 1.9991 & $-1.5$-$0.8(-1.9$-$2.4)$ & $-3.2$-$-2.2(-3.6$-$-1.6)$ & $1.3$-$3.6(1.0$-$4.3)$ & $ 0.3$-$3.6(-0.1$-$5.3)$ & \\
3 & Q0458$-$02     & 2.0395 & $1.6$-$2.7(1.3$-$2.9)$ & $-2.8$-$-1.3(-3.1$-$-0.9)$ & $1.7$-$2.5(1.5$-$3.2)$ & $ 3.6$-$4.9( 3.2$-$5.6)$ & \\
4 & J2340$-$0053   & 2.0545 & $1.0$-$2.2(0.8$-$2.8)$ & $-2.9$-$-1.5(-3.2$-$-1.1)$ & $1.6$-$2.5(1.4$-$3.2)$ & $ 2.8$-$4.3( 2.5$-$5.2)$ & \\
5 & Q2206$-$19     & 2.0762 & $-1.2$-$1.3(-1.8$-$2.6)$ & $-3.0$-$-1.8(-3.5$-$-1.3)$ & $1.4$-$3.7(1.1$-$4.3)$ & $ 0.9$-$4.3( 0.1$-$5.6)$ & \\
6 & Q2359$-$02     & 2.0950 & $0.3$-$1.5(0.0$-$2.7)$ & $-2.7$-$-1.6(-3.1$-$-1.3)$ & $1.7$-$3.8(1.3$-$4.3)$ & $ 2.3$-$4.7( 1.8$-$5.9)$ & \\
7 & Q0149$+$33     & 2.1407 & $-1.4$-$0.9(-1.9$-$2.4)$ & $-3.2$-$-2.1(-3.5$-$-1.6)$ & $1.3$-$3.6(1.1$-$4.3)$ & $ 0.4$-$3.7(-0.1$-$5.3)$ & \\
8 & Q2348$-$14     & 2.2794 & $-1.3$-$1.1(-1.9$-$2.5)$ & $-3.1$-$-2.0(-3.5$-$-1.4)$ & $1.3$-$3.7(1.1$-$4.3)$ & $ 0.7$-$4.0( 0.0$-$5.5)$ & \\
9 & J2036$-$0553   & 2.2805 & $0.6$-$1.7(0.3$-$2.7)$ & $-2.7$-$-1.5(-3.1$-$-1.1)$ & $1.7$-$3.8(1.4$-$4.4)$ & $ 2.5$-$5.0( 2.1$-$6.0)$ & \\
10 & J1435$+$5359   & 2.3427 & $-0.3$-$1.5(-0.6$-$2.7)$ & $-2.7$-$-1.8(-3.1$-$-1.3)$ & $1.5$-$3.7(1.2$-$4.3)$ & $ 1.6$-$4.5( 1.2$-$5.8)$ & \\
11 & Q2343$+$125    & 2.4313 & $-1.3$-$0.8(-1.8$-$2.5)$ & $-3.0$-$-2.2(-3.3$-$-1.6)$ & $1.5$-$3.8(1.1$-$4.3)$ & $ 0.6$-$3.7( 0.1$-$5.5)$ & \\
12 & J1541$+$3153   & 2.4435 & $0.1$-$1.4(-0.3$-$2.6)$ & $-2.7$-$-1.7(-3.1$-$-1.3)$ & $1.7$-$3.9(1.3$-$4.4)$ & $ 2.1$-$4.7( 1.6$-$5.9)$ & \\
13 & Q0836$+$11     & 2.4652 & $-1.5$-$0.9(-1.9$-$2.4)$ & $-3.2$-$-2.2(-3.6$-$-1.6)$ & $1.3$-$3.6(1.0$-$4.3)$ & $ 0.3$-$3.6(-0.2$-$5.2)$ & \\
14 & Q2344$+$12     & 2.5379 & $-1.3$-$1.0(-1.9$-$2.5)$ & $-3.1$-$-2.0(-3.5$-$-1.4)$ & $1.3$-$3.7(1.1$-$4.3)$ & $ 0.6$-$3.9( 0.0$-$5.5)$ & \\
15 & Q0913$+$072    & 2.6184 & $-1.1$-$1.5(-1.8$-$2.6)$ & $-3.0$-$-1.7(-3.4$-$-1.1)$ & $1.4$-$3.7(1.1$-$4.3)$ & $ 1.1$-$4.5( 0.2$-$5.7)$ & \\
16 & Q1759$+$75     & 2.6253 & $-0.6$-$1.0(-0.9$-$2.5)$ & $-2.8$-$-2.0(-3.1$-$-1.5)$ & $1.6$-$3.8(1.2$-$4.3)$ & $ 1.3$-$4.1( 0.8$-$5.6)$ & \\
17 & J1035$+$5440   & 2.6840 & $-1.0$-$1.9(-1.8$-$2.8)$ & $-3.0$-$-1.7(-3.4$-$-1.1)$ & $1.5$-$3.6(1.1$-$4.3)$ & $ 1.1$-$4.7( 0.2$-$5.8)$ & \\
18 & PKS1354$-$17   & 2.7800 & $0.2$-$1.4(-0.6$-$2.6)$ & $-2.7$-$-1.6(-3.1$-$-1.3)$ & $1.7$-$3.8(1.3$-$4.4)$ & $ 2.2$-$4.7( 1.5$-$5.9)$ & \\
19 & HS1132$+$2243  & 2.7834 & $-1.4$-$0.8(-1.9$-$2.4)$ & $-3.1$-$-2.1(-3.5$-$-1.6)$ & $1.3$-$3.7(1.1$-$4.3)$ & $ 0.5$-$3.7( 0.0$-$5.3)$ & \\
20 & Q1337$+$11     & 2.7958 & $-1.5$-$0.9(-1.9$-$2.4)$ & $-3.2$-$-2.2(-3.6$-$-1.6)$ & $1.2$-$3.6(1.0$-$4.3)$ & $ 0.3$-$3.6(-0.2$-$5.2)$ & \\
21 & J1353$+$5328   & 2.8349 & $0.5$-$1.6(0.2$-$2.7)$ & $-2.7$-$-1.5(-3.1$-$-1.2)$ & $1.7$-$3.8(1.4$-$4.3)$ & $ 2.5$-$4.9( 2.0$-$6.0)$ & \\
22 & J1131$+$6044   & 2.8757 & $-1.3$-$1.1(-1.8$-$2.5)$ & $-3.1$-$-1.9(-3.5$-$-1.4)$ & $1.4$-$3.7(1.1$-$4.3)$ & $ 0.7$-$4.0( 0.1$-$5.5)$ & \\
23 & J1304$+$1202   & 2.9131 & $-1.3$-$1.0(-1.9$-$2.4)$ & $-3.1$-$-2.0(-3.5$-$-1.5)$ & $1.3$-$3.7(1.1$-$4.3)$ & $ 0.6$-$3.9( 0.0$-$5.4)$ & \\
24 & J1304$+$1202   & 2.9289 & $-1.1$-$1.2(-1.8$-$2.5)$ & $-3.0$-$-1.8(-3.5$-$-1.3)$ & $1.4$-$3.7(1.1$-$4.3)$ & $ 0.9$-$4.3( 0.2$-$5.6)$ & \\
25 & Q1021$+$30     & 2.9489 & $-1.4$-$0.9(-1.9$-$2.4)$ & $-3.1$-$-2.1(-3.5$-$-1.5)$ & $1.3$-$3.6(1.1$-$4.3)$ & $ 0.4$-$3.7(-0.1$-$5.3)$ & \\
26 & J1014$+$4300   & 2.9588 & $-1.4$-$0.9(-1.9$-$2.4)$ & $-3.1$-$-2.1(-3.5$-$-1.6)$ & $1.3$-$3.6(1.1$-$4.3)$ & $ 0.4$-$3.7(-0.1$-$5.3)$ & \\
27 & J1410$+$5111   & 2.9642 & $-1.4$-$1.0(-1.9$-$2.5)$ & $-3.1$-$-2.0(-3.5$-$-1.5)$ & $1.3$-$3.7(1.1$-$4.3)$ & $ 0.5$-$3.9( 0.0$-$5.4)$ & \\
28 & HS0741$+$4741  & 3.0174 & $-1.5$-$0.9(-1.9$-$2.4)$ & $-3.2$-$-2.2(-3.6$-$-1.6)$ & $1.2$-$3.5(1.0$-$4.3)$ & $ 0.2$-$3.5(-0.2$-$5.2)$ & \\
29 & J1240$+$1455   & 3.0241 & $-1.1$-$1.2(-1.8$-$2.5)$ & $-3.0$-$-1.8(-3.5$-$-1.3)$ & $1.4$-$3.8(1.1$-$4.3)$ & $ 1.0$-$4.3( 0.2$-$5.6)$ & \\
30 & Q0347$-$38     & 3.0247 & $0.6$-$1.7(0.3$-$2.7)$ & $-2.7$-$-1.5(-3.1$-$-1.2)$ & $1.7$-$3.8(1.4$-$4.4)$ & $ 2.6$-$5.0( 2.2$-$6.0)$ & \\
31 & Q0336$-$01     & 3.0620 & $-1.0$-$1.6(-1.8$-$2.6)$ & $-3.0$-$-1.7(-3.4$-$-1.1)$ & $1.4$-$3.7(1.1$-$4.3)$ & $ 1.1$-$4.6( 0.3$-$5.8)$ & \\
32 & J1200$+$4015   & 3.2200 & $-1.2$-$1.2(-1.8$-$2.5)$ & $-3.1$-$-1.9(-3.5$-$-1.4)$ & $1.4$-$3.7(1.1$-$4.3)$ & $ 0.8$-$4.1( 0.1$-$5.5)$ & \\
33 & Q0930$+$28     & 3.2352 & $-1.3$-$1.0(-1.9$-$2.5)$ & $-3.1$-$-2.0(-3.5$-$-1.4)$ & $1.3$-$3.7(1.1$-$4.3)$ & $ 0.6$-$3.9( 0.0$-$5.5)$ & \\
34 & J0900$+$4215   & 3.2458 & $-0.3$-$1.2(-0.6$-$2.7)$ & $-2.8$-$-1.8(-3.1$-$-1.4)$ & $1.6$-$3.7(1.2$-$4.3)$ & $ 1.7$-$4.3( 1.2$-$5.8)$ & \\
35 & J0929$+$2825   & 3.2627 & $0.2$-$1.5(-0.1$-$2.7)$ & $-2.8$-$-1.7(-3.1$-$-1.4)$ & $1.6$-$3.6(1.3$-$4.3)$ & $ 2.1$-$4.5( 1.7$-$5.8)$ & \\
36 & J2315$+$1456   & 3.2732 & $-1.3$-$1.1(-1.9$-$2.5)$ & $-3.1$-$-2.0(-3.5$-$-1.4)$ & $1.3$-$3.7(1.1$-$4.3)$ & $ 0.7$-$4.0( 0.0$-$5.5)$ & \\
37 & BR0019$-$15    & 3.4388 & $0.5$-$1.7(0.2$-$2.7)$ & $-2.7$-$-1.5(-3.1$-$-1.2)$ & $1.7$-$3.8(1.4$-$4.4)$ & $ 2.4$-$4.9( 2.0$-$6.0)$ & \\
38 & J0814$+$5029   & 3.7075 & $0.7$-$1.7(0.4$-$2.7)$ & $-2.7$-$-1.4(-3.1$-$-1.1)$ & $1.8$-$3.9(1.4$-$4.4)$ & $ 2.8$-$5.1( 2.3$-$6.1)$ & \\
39 & BRI1346$-$03   & 3.7358 & $0.4$-$1.6(-0.8$-$2.7)$ & $-2.7$-$-1.5(-3.1$-$-1.1)$ & $1.7$-$3.9(1.4$-$4.4)$ & $ 2.5$-$4.9( 1.4$-$6.0)$ & \\
40 & PSS0209$+$05   & 3.8635 & $-1.3$-$1.0(-1.9$-$2.5)$ & $-3.1$-$-2.0(-3.5$-$-1.4)$ & $1.4$-$3.7(1.1$-$4.3)$ & $ 0.7$-$3.9( 0.1$-$5.5)$ & \\
41 & J1051$+$3107   & 4.1392 & $-1.3$-$1.1(-1.9$-$2.5)$ & $-3.1$-$-2.0(-3.5$-$-1.4)$ & $1.3$-$3.7(1.1$-$4.3)$ & $ 0.6$-$4.0( 0.0$-$5.5)$ & \\
42 & PSS1443$+$27   & 4.2241 & $-1.1$-$1.6(-1.8$-$2.7)$ & $-3.0$-$-1.8(-3.5$-$-1.3)$ & $1.4$-$3.5(1.1$-$4.3)$ & $ 0.9$-$4.4( 0.1$-$5.7)$ & \\
43 & J0817$+$1351   & 4.2584 & $-0.9$-$2.0(-1.8$-$2.8)$ & $-3.0$-$-1.6(-3.4$-$-1.0)$ & $1.5$-$3.7(1.1$-$4.3)$ & $ 1.3$-$4.9( 0.3$-$5.9)$ & \\
44 & J1100$+$1122   & 4.3947 & $-1.0$-$1.7(-1.8$-$2.7)$ & $-3.0$-$-1.6(-3.4$-$-1.0)$ & $1.5$-$3.7(1.1$-$4.3)$ & $ 1.2$-$4.8( 0.3$-$5.9)$ & \\
45 & J1607$+$1604   & 4.4741 & $1.0$-$1.9(0.6$-$2.7)$ & $-2.6$-$-1.2(-3.1$-$-0.9)$ & $1.9$-$3.9(1.5$-$4.4)$ & $ 3.2$-$5.4( 2.7$-$6.3)$ & \\
46 & J1200$+$4618   & 4.4765 & $-1.1$-$1.3(-1.8$-$2.6)$ & $-3.0$-$-1.7(-3.4$-$-1.2)$ & $1.4$-$3.7(1.1$-$4.3)$ & $ 1.0$-$4.4( 0.2$-$5.7)$ & \\
47 & J1202$+$3235   & 4.7955 & $-0.9$-$1.8(-1.8$-$2.6)$ & $-3.0$-$-1.6(-3.4$-$-1.0)$ & $1.5$-$3.8(1.1$-$4.4)$ & $ 1.4$-$4.9( 0.4$-$5.9)$ & \\
48 & J1051$+$3545   & 4.8206 & $-1.2$-$1.3(-1.8$-$2.5)$ & $-3.0$-$-1.8(-3.5$-$-1.2)$ & $1.4$-$3.7(1.1$-$4.3)$ & $ 0.9$-$4.3( 0.2$-$5.6)$ & \\
49 & J1056$+$1208   & 1.6093 & $1.6$-$2.5(1.5$-$2.9)$ & $-2.7$-$-1.2(-3.0$-$-0.9)$ & $2.2$-$2.8(2.1$-$3.2)$ & $ 4.0$-$5.1( 3.9$-$5.6)$ & \\
50 & J0044$+$0018   & 1.7250 & $0.6$-$1.7(0.4$-$2.7)$ & $-2.7$-$-1.4(-3.1$-$-1.1)$ & $1.8$-$3.8(1.4$-$4.4)$ & $ 2.6$-$5.0( 2.2$-$6.1)$ & \\
51 & J0927$+$1543   & 1.7311 & $2.0$-$2.7(1.8$-$2.9)$ & $-2.8$-$-1.2(-3.1$-$-0.9)$ & $1.7$-$2.1(1.6$-$2.3)$ & $ 3.8$-$4.6( 3.6$-$5.0)$ & \\
52 & J0008$-$0958   & 1.7675 & $0.8$-$2.3(0.6$-$2.8)$ & $-2.9$-$-1.6(-3.2$-$-1.1)$ & $1.5$-$2.4(1.3$-$2.9)$ & $ 2.6$-$4.3( 2.3$-$5.0)$ & \\
53 & J1249$-$0233   & 1.7808 & $-1.1$-$1.4(-1.8$-$2.6)$ & $-3.0$-$-1.7(-3.4$-$-1.2)$ & $1.5$-$3.8(1.1$-$4.3)$ & $ 1.1$-$4.5( 0.3$-$5.7)$ & \\
54 & J0233$+$0103   & 1.7850 & $-1.4$-$1.0(-1.9$-$2.4)$ & $-3.2$-$-2.1(-3.6$-$-1.5)$ & $1.3$-$3.6(1.0$-$4.3)$ & $ 0.4$-$3.7(-0.1$-$5.3)$ & \\
55 & J1454$+$0941   & 1.7884 & $0.2$-$1.4(-0.1$-$2.6)$ & $-2.7$-$-1.6(-3.1$-$-1.3)$ & $1.7$-$3.9(1.3$-$4.4)$ & $ 2.2$-$4.7( 1.7$-$5.9)$ & \\
56 & J1313$+$1441   & 1.7947 & $1.7$-$2.5(1.6$-$2.9)$ & $-2.7$-$-1.2(-3.0$-$-0.9)$ & $2.3$-$2.8(2.2$-$3.1)$ & $ 4.1$-$5.2( 4.0$-$5.6)$ & \\
57 & J1310$+$5424   & 1.8005 & $1.6$-$2.6(1.4$-$2.9)$ & $-2.8$-$-1.3(-3.1$-$-0.9)$ & $2.1$-$2.6(2.0$-$3.0)$ & $ 3.9$-$5.0( 3.6$-$5.4)$ & \\
58 & J1106$+$1044   & 1.8185 & $-1.4$-$0.9(-1.9$-$2.4)$ & $-3.1$-$-2.1(-3.5$-$-1.6)$ & $1.3$-$3.6(1.1$-$4.3)$ & $ 0.5$-$3.7(-0.1$-$5.3)$ & \\
59 & J1142$+$0701   & 1.8407 & $0.9$-$2.2(0.6$-$2.8)$ & $-2.6$-$-1.2(-3.0$-$-0.9)$ & $2.3$-$3.9(2.0$-$4.4)$ & $ 3.6$-$5.6( 3.1$-$6.4)$ & \\
60 & J0815$+$1037   & 1.8462 & $-1.4$-$0.9(-1.9$-$2.4)$ & $-3.1$-$-2.1(-3.5$-$-1.5)$ & $1.3$-$3.6(1.1$-$4.3)$ & $ 0.4$-$3.7(-0.1$-$5.3)$ & \\
61 & J1335$+$0824   & 1.8560 & $0.2$-$1.4(0.0$-$2.7)$ & $-2.7$-$-1.6(-3.1$-$-1.3)$ & $1.7$-$3.8(1.3$-$4.3)$ & $ 2.2$-$4.7( 1.8$-$5.9)$ & \\
62 & J1024$+$0600   & 1.8950 & $0.9$-$2.1(0.7$-$2.8)$ & $-2.8$-$-1.6(-3.2$-$-1.1)$ & $1.6$-$2.9(1.4$-$3.8)$ & $ 2.8$-$4.5( 2.4$-$5.6)$ & \\
63 & J1524$+$1030   & 1.9409 & $-1.3$-$2.0(-1.9$-$2.8)$ & $-3.1$-$-2.0(-3.5$-$-1.2)$ & $1.3$-$3.1(1.1$-$4.2)$ & $ 0.4$-$4.2(-0.2$-$5.5)$ & \\
64 & J1042$+$0628   & 1.9429 & $-1.4$-$1.0(-1.9$-$2.5)$ & $-3.1$-$-2.0(-3.5$-$-1.5)$ & $1.3$-$3.6(1.1$-$4.3)$ & $ 0.5$-$3.8( 0.0$-$5.4)$ & \\
65 & J1417$+$4132   & 1.9508 & $-1.0$-$1.9(-1.8$-$2.8)$ & $-3.0$-$-1.8(-3.4$-$-1.2)$ & $1.4$-$3.5(1.1$-$4.3)$ & $ 1.0$-$4.6( 0.1$-$5.8)$ & \\
66 & J1552$+$4910   & 1.9599 & $-0.5$-$1.4(-0.9$-$2.7)$ & $-2.9$-$-2.0(-3.2$-$-1.4)$ & $1.5$-$3.4(1.2$-$4.2)$ & $ 1.3$-$4.0( 0.8$-$5.5)$ & \\
67 & Q1755$+$578    & 1.9692 & $0.7$-$2.0(0.5$-$2.7)$ & $-2.5$-$-1.2(-2.9$-$-1.0)$ & $2.5$-$4.0(2.2$-$4.4)$ & $ 3.5$-$5.5( 3.2$-$6.3)$ & \\
68 & J1305$+$0924   & 2.0184 & $0.6$-$1.9(0.3$-$2.8)$ & $-2.8$-$-1.6(-3.1$-$-1.2)$ & $1.6$-$3.4(1.4$-$4.2)$ & $ 2.5$-$4.7( 2.1$-$5.9)$ & \\
69 & J1509$+$1113   & 2.0283 & $0.7$-$2.2(0.4$-$2.8)$ & $-2.9$-$-1.7(-3.2$-$-1.2)$ & $1.5$-$2.5(1.3$-$3.3)$ & $ 2.5$-$4.3( 2.2$-$5.3)$ & \\
70 & J1135$-$0010   & 2.2068 & $1.7$-$2.6(1.5$-$2.9)$ & $-2.5$-$-0.8(-2.8$-$-0.6)$ & $2.8$-$4.1(2.6$-$4.4)$ & $ 4.8$-$6.4( 4.5$-$6.8)$ & \\
71 & J1211$+$0422   & 2.3765 & $-1.6$-$0.8(-1.9$-$2.3)$ & $-3.2$-$-2.3(-3.6$-$-1.6)$ & $1.2$-$3.4(1.0$-$4.3)$ & $ 0.0$-$3.3(-0.3$-$5.0)$ & \\
72 & J2241$+$1225   & 2.4179 & $-1.1$-$1.3(-1.8$-$2.5)$ & $-3.0$-$-1.8(-3.5$-$-1.2)$ & $1.4$-$3.7(1.1$-$4.3)$ & $ 1.0$-$4.4( 0.2$-$5.7)$ & \\
73 & J0211$+$1241   & 2.5951 & $-0.4$-$1.1(-0.7$-$2.5)$ & $-2.8$-$-1.9(-3.1$-$-1.5)$ & $1.6$-$3.8(1.3$-$4.4)$ & $ 1.6$-$4.3( 1.1$-$5.7)$ & \\
74 & J0812$+$3208   & 2.6263 & $0.6$-$2.2(0.2$-$2.8)$ & $-2.9$-$-1.8(-3.2$-$-1.2)$ & $1.4$-$2.3(1.3$-$2.7)$ & $ 2.2$-$4.1( 1.8$-$4.9)$ & \\
75 & J1558$-$0031   & 2.7026 & $-1.3$-$1.0(-1.8$-$2.5)$ & $-3.1$-$-2.0(-3.5$-$-1.4)$ & $1.3$-$3.7(1.1$-$4.3)$ & $ 0.7$-$4.0( 0.1$-$5.5)$ & \\
76 & FJ2334$-$09    & 3.0569 & $-1.6$-$0.9(-1.9$-$2.4)$ & $-3.3$-$-2.4(-3.7$-$-1.6)$ & $1.1$-$3.0(1.0$-$4.2)$ & $-0.2$-$3.0(-0.5$-$4.9)$ & \\
77 & J2100$-$0641   & 3.0924 & $0.8$-$2.3(0.5$-$2.8)$ & $-2.9$-$-1.7(-3.2$-$-1.1)$ & $1.4$-$2.2(1.3$-$2.5)$ & $ 2.5$-$4.1( 2.1$-$4.8)$ & \\
78 & J1155$+$0530   & 3.3260 & $-0.2$-$1.7(-0.7$-$2.7)$ & $-2.9$-$-2.0(-3.2$-$-1.3)$ & $1.4$-$2.6(1.2$-$3.8)$ & $ 1.5$-$3.8( 1.0$-$5.2)$ & \\
79 & J0825$+$3544   & 3.6567 & $-0.3$-$1.5(-1.4$-$2.6)$ & $-2.8$-$-1.8(-3.2$-$-1.4)$ & $1.5$-$3.6(1.2$-$4.3)$ & $ 1.7$-$4.4( 0.6$-$5.7)$ & \\
80 & J0909$+$3303   & 3.6581 & $0.6$-$1.7(-0.3$-$2.7)$ & $-2.7$-$-1.4(-3.1$-$-1.1)$ & $1.8$-$3.9(1.4$-$4.4)$ & $ 2.7$-$5.1( 1.9$-$6.1)$ & 
\enddata
\end{deluxetable*}

\end{document}